\documentclass[gmd, manuscript]{copernicus}
\usepackage{changes}

\begin{document}
\nolinenumbers
\title{ADEPTS: An auto-differentiable framework for time-dependent nonlinear thermo-chemical mantle convection inversion}


\Author[1]{Zhiying}{Ming}
\Author[1,2,*][hujs@sustech.edu.cn]{Jiashun}{Hu}

\affil[1]{Department of Earth and Space Sciences, Southern University of Science and Technology, Shenzhen, China, 518055}
\affil[2]{Guangdong Provincial Key Laboratory of Geophysical High-resolution Imaging Technology, Southern University of Science and Technology, Shenzhen, China, 518055}




\runningtitle{ADEPTS: Differentiable mantle convection inversion}
\runningauthor{Ming and Hu}

\received{}
\pubdiscuss{} 
\revised{}
\accepted{}
\published{}


\firstpage{1}

\maketitle

\graphicspath{{figure/}}

\begin{abstract}
Time-dependent mantle-dynamics inversion must simultaneously address the high dimensionality of the initial state, nonlinear rheology, and the complex propagation of gradients through long-term thermo-mechanical evolution.
Here, we develop a geodynamic inversion framework \textit{ADEPTS} based on a two-dimensional staggered-grid finite-difference discretization and automatic differentiation, with the aim of jointly constraining the initial temperature field and mantle material properties.
The forward model solves the incompressible Stokes equations, the temperature advection--diffusion equation, and the compositional advection equation, while accounting for temperature- and strain-rate-dependent nonlinear viscosity and plastic yielding.
For the nonlinear Stokes system arising at each time step, we implement two gradient-computation strategies: unrolled differentiation through a fixed number of Picard iterations and implicit differentiation based on the sufficiently converged discrete nonlinear residual equations.
Numerical experiments show that 
unrolled differentiation 
enables stable inversion even when the nonlinear solve has not fully converged, 
while the accuracy of implicit differentiation depends on sufficient convergence of the nonlinear Stokes system.
When the nonlinear solve is insufficiently accurate, both gradient consistency and optimization convergence deteriorate for implicit differentiation, whereas sufficiently converged nonlinear solves recover the expected gradient accuracy and yield reconstruction results comparable to those obtained with unrolled differentiation.
Under idealized noise-free twin-experiment conditions, we further test the joint inversion of the initial temperature field and physical parameters. The joint inversion recovers the initial temperature field together with the density of a compositional anomaly, the reference viscosity, and the stress exponent, while simultaneously fitting observations of the final-time temperature field, surface horizontal velocity, and surface normal stress.
The two differentiation strategies impose different requirements on the accuracy of the nonlinear Stokes solve, but both can be used for gradient computation and inversion within a differentiable time-dependent mantle-dynamics framework.
\end{abstract}


\introduction
\label{sec:introduction}

Mantle convection is an inherently time-dependent dynamical process in which the present-day state of the Earth's interior reflects the cumulative evolution of earlier thermal and compositional structures under uncertain material properties.
The initial distribution of temperature and composition influences the subsequent evolution of mantle flow, while parameters such as viscosity and density control buoyancy, deformation, and the characteristic timescales of convection
\citep{davies1999dynamic,ranalli1995rheology,karato2008deformation}.
These factors jointly affect the morphology and penetration of subducted slabs
\citep{garel2014interaction,rudolph2015viscosity,goes2017subduction,li2019variability,fei2023variation,li2026dual},
the evolution of thermo-chemical structures in the deep mantle
\citep{mcnamara2005thermochemical,torsvik2008long,zhang2024segregation,shi2024sluggish},
and surface expressions of mantle dynamics, including plate motions, dynamic topography, and the geoid
\citep{hu2024constraining,becker2001predicting,hager1985lower,steinberger2006models}.
Reconstructing mantle evolution thus requires more than determining a set of present-day material parameters:
it also requires inference of the earlier mantle state from which the observed system evolved.
This reconstruction is fundamentally an inverse problem.
Forward geodynamic models can predict mantle evolution for prescribed initial conditions and material parameters, but the thermal and compositional structure of the mantle in the geological past is generally not directly observable.
Available constraints instead come from present-day seismic structure, plate-motion histories, geological reconstructions, and surface responses that record aspects of the preceding dynamical evolution
\citep{seton2012global,bunge2003mantle,liu2008simultaneous,spasojevic2009adjoint}.
The problem is formulated as a time-dependent inverse problem in which an initial mantle state and a set of physical parameters are propagated forward in time and iteratively adjusted to reduce the discrepancy between model predictions and observations.
This formulation differs fundamentally from instantaneous or steady-state parameter inversion because errors in the initial state and material parameters are propagated through the entire dynamical trajectory before reaching the observations.

Variational data assimilation and adjoint methods have provided an important framework for solving such time-dependent mantle inverse problems.
\citet{bunge2003mantle} showed that straightforward backward integration is unable to reconstruct past mantle structure reliably because thermal diffusion introduces irreversible information loss, and demonstrated that adjoint optimization constrained by forward mantle convection provides a more physically consistent alternative.
\citet{liu2008simultaneous} subsequently extended this approach to the simultaneous inversion of mantle initial conditions and material properties.
These two studies assumed linear mantle rheology and used present-day mantle structure as the main constraint.
Building on these previous works, \citet{li2017towards} developed a time-dependent adjoint formulation with temperature- and strain-rate-dependent viscosity and demonstrated joint inversion of a high-dimensional initial temperature field and low-dimensional viscosity parameters in synthetic subduction models, with constraints from both present-day mantle structure and historical plate motion.
More recently, \citet{nakao2024adjoint} investigated how final-time thermal structure and time-dependent surface observations constrain different spatial scales of mantle convection, while \citet{ghelichkhan2024automatic} introduced the G-ADOPT framework for automated adjoint inversion of time-dependent mantle convection in both isoviscous and nonlinear viscoplastic settings.
These studies demonstrate that mantle histories can be reconstructed through gradient-based optimization and highlight the need for accurate and computationally tractable propagation of sensitivities through long-term nonlinear thermo-mechanical evolution.

Automatic differentiation provides a flexible alternative for building differentiable geodynamic models by representing the discrete numerical algorithm as a computational graph and propagating derivatives through the underlying numerical operations
\citep{sambridge2007automatic,griewank2008evaluating,baydin2018automatic,ming2026application}.
For scalar objective functions with high-dimensional control variables, reverse-mode automatic differentiation is particularly well suited to gradient-based inverse problems
\citep{ming2026application}.
A key challenge arises from the nonlinear Stokes system that must be solved at every time step.
Unlike the explicit parts of the time-stepping scheme, the Stokes solution is obtained iteratively and thus needs special treatment when gradients are computed.
Two main strategies can be used.
One is to differentiate through the finite sequence of nonlinear iterations performed by the forward solver, while the other is to treat the sufficiently converged Stokes solution as an implicit function of the nonlinear residual equations.
The first approach gives the gradient of the finite-iteration forward model actually used in the calculation.
The second avoids differentiating through the full nonlinear iteration history, but relies on the Stokes solution being sufficiently converged.
As a result, the two strategies may have different requirements in terms of nonlinear convergence, memory use, and computational cost.
These differences become especially important in time-dependent inversion, where the nonlinear Stokes problem has to be solved repeatedly throughout the forward evolution.

Compared with traditional adjoint methods, automatic differentiation provides a more flexible way to incorporate multiple observational constraints and different types of inversion variables into geodynamic inversion
\citep{ming2026application}.
Laboratory deformation experiments provide important constraints on mantle rheology
\citep{karato1993rheology,hirth2003rheology,korenaga2008new,jackson2015},
but extrapolating these results to geological strain rates and mantle pressure--temperature conditions remains uncertain.
Geodynamic observations, including plate velocities, stress orientations, gravity and geoid anomalies, and surface deformation, can provide additional constraints on mantle material properties
\citep{forte1991inferences,becker2001predicting,Reuber2020,liu2015long,hu2024constraining,ming2026application}.
In time-dependent inversion, uncertainties in the initial thermal structure can also affect the inferred density and rheological parameters, while uncertainties in compositional buoyancy may be partly compensated by changes in the recovered temperature field.
These considerations make joint inversion of high-dimensional initial conditions and low-dimensional physical parameters useful not only for estimating model parameters, but also for exploring trade-offs between different sources of mantle buoyancy and deformation.

In this study, we develop a two-dimensional time-dependent thermo-mechanical inversion framework that combines staggered-grid finite differences with automatic differentiation.
PyTorch is used to handle the automatic differentiation
\citep{paszke2019pytorch}.
Temperature diffusion and advection, compositional advection, staggered-grid interpolation, observation operators, and objective-function evaluation are all written using differentiable tensor operations, so their derivatives can be obtained directly through reverse-mode differentiation.
This makes it possible to differentiate most of the time-dependent forward model without deriving a separate adjoint equation for each numerical component.
The forward model couples incompressible Stokes flow with temperature advection--diffusion and compositional advection, and includes temperature- and strain-rate-dependent viscosity as well as plastic yielding.
The nonlinear Stokes solve requires separate treatment, for which we consider two approaches: unrolled differentiation through a prescribed number of Picard iterations and implicit differentiation based on sufficiently converged nonlinear residual equations.
We conduct a series of synthetic experiments to compare these two approaches and assess how well the framework can recover both high-dimensional initial conditions and low-dimensional physical parameters, including rheological and compositional parameters, from different types of observations.

\section{Methods}

\subsection{Forward model}

This section presents the forward model used to describe the thermo-mechanical evolution of the mantle and to 
support the subsequent inversion. The model is based on incompressible Stokes flow, 
coupled with temperature and compositional transport, and incorporates nonlinear viscous rheology. 
The following subsections describe the governing equations, staggered-grid finite-difference discretization, 
nonlinear solution strategy, and the time-integration schemes for the temperature and compositional fields.

\subsubsection{Governing equations}
\label{subsubsec:governing_equations}

We consider a two-dimensional thermo-mechanical model under the incompressible Boussinesq approximation to describe the slow deformation of the mantle governed by creeping viscous flow and the associated surface dynamic response.
The model is defined in a two-dimensional Cartesian domain $\Omega$.
The velocity field is denoted by $\mathbf{u}=(u_x,u_z)$, the pressure by $p$, and the temperature by $T$.
In some thermo-chemical experiments, a compositional field $C$ is additionally introduced to represent buoyancy arising from compositional density contrasts.

Conservation of mass for an incompressible fluid is given by
\begin{equation}
\nabla\cdot\mathbf{u}=0.
\label{eq:mass_conservation}
\end{equation}

Under the infinite-Prandtl-number approximation, in which inertial effects are neglected, 
the momentum-conservation equation is written as
\begin{equation}
-\nabla\cdot\boldsymbol{\sigma}
=
\rho(T,C)\mathbf{g},
\label{eq:momentum_conservation}
\end{equation}
where $\boldsymbol{\sigma}$ is the stress tensor,
$\mathbf{g}$ is the gravitational acceleration vector,
and $\rho(T,C)$ is the density accounting for thermal and compositional effects.
Under the Boussinesq approximation, density variations are retained only in the gravitational body-force term.

The stress tensor is defined as
\begin{equation}
\boldsymbol{\sigma}
=
-p\mathbf{I}
+
2\eta_{\mathrm{eff}}
\dot{\boldsymbol{\varepsilon}}(\mathbf{u}),
\label{eq:stress_tensor}
\end{equation}
where $\eta_{\mathrm{eff}}$ is the effective viscosity,
$\mathbf{I}$ is the identity tensor,
and $\dot{\boldsymbol{\varepsilon}}(\mathbf{u})$ is the strain-rate tensor, defined as
\begin{equation}
\dot{\boldsymbol{\varepsilon}}(\mathbf{u})
=
\frac{1}{2}
\left[
\nabla\mathbf{u}
+
\left(\nabla\mathbf{u}\right)^{T}
\right].
\label{eq:strain_rate_tensor}
\end{equation}

Neglecting internal heating, adiabatic heating, and viscous dissipation,
the temperature field satisfies the advection--diffusion equation
\begin{equation}
\rho_0 c_p
\left(
\frac{\partial T}{\partial t}
+
\mathbf{u}\cdot\nabla T
\right)
=
\nabla\cdot
\left(
k\nabla T
\right),
\label{eq:temperature_equation}
\end{equation}
where $c_p$ is the specific heat capacity at constant pressure and $k$ is the thermal conductivity.

For simulations that include compositional heterogeneity,
the compositional field is treated as a non-diffusive scalar advected by the velocity field:
\begin{equation}
\frac{\partial C}{\partial t}
+
\mathbf{u}\cdot\nabla C
=
0.
\label{eq:composition_equation}
\end{equation}

\subsubsection{Staggered-grid finite-difference discretization}
\label{subsubsec:staggered_grid}

The spatial discretization is based on a two-dimensional staggered-grid finite-difference scheme, following the variable arrangement commonly used in numerical geodynamics \citep{gerya2019introduction}.
Pressure $p$, temperature $T$, and composition $C$ are defined at cell centres on the $p$-grid.
The horizontal velocity component $u_x$ is defined on the left and right cell faces, whereas the vertical velocity component $u_z$ is defined on the lower and upper cell faces. Effective viscosity is evaluated at cell centres for the normal-stress components and at grid nodes for the shear-stress components.

The pressure gradient, velocity divergence, and stress divergence are discretized using local finite-difference stencils.
For non-uniform grids, the discretization coefficients are computed using the actual distances between neighbouring grid points.
Viscosity values required at different variable locations are obtained through the corresponding grid interpolation operators.
The discrete velocity and pressure unknowns are assembled into a unified Stokes state vector,
\begin{equation}
\mathbf{s}
=
\begin{bmatrix}
\mathbf{u}_x &
\mathbf{u}_z &
\mathbf{p}
\end{bmatrix}^{T}.
\label{eq:stokes_state_vector}
\end{equation}

Depending on the numerical experiment, the velocity boundary conditions are specified as either free-slip or no-slip.
The free-slip condition is written as
\begin{equation}
\mathbf{u}\cdot\mathbf{n}=0,
\qquad
\mathbf{t}\cdot
\boldsymbol{\sigma}\mathbf{n}=0,
\label{eq:free_slip_boundary}
\end{equation}
where $\mathbf{n}$ and $\mathbf{t}$ denote the unit normal and tangential vectors to the boundary, respectively.
The temperature field supports both Dirichlet and zero-flux Neumann boundary conditions, whereas the compositional field is typically subject to zero-flux boundary conditions.
Because the pressure in the incompressible Stokes equations is determined only up to an arbitrary additive constant, one pressure degree of freedom is fixed to remove the pressure null space.

The staggered-grid Stokes discretization, boundary-condition treatment, and linear-system assembly used in this study have been verified previously against steady-state Stokes benchmarks \citep{ming2026application}.
Accordingly, the present study focuses on validating the newly introduced time-integration, nonlinear implicit-differentiation, and thermo-chemical inversion components.

\subsubsection{Nonlinear viscosity and nonlinear solution}
\label{subsubsec:nonlinear_viscosity_solver}

We employ a dislocation-creep rheology with temperature- and strain-rate-dependent viscosity and, in selected experiments, additionally account for plastic yielding.
The dislocation-creep viscosity is written as
\begin{equation}
\eta_{\mathrm{disl}}
=
\eta_{\mathrm{ref}}
\left(
\frac{
\dot{\varepsilon}_{II}
}{
\dot{\varepsilon}_{\mathrm{ref}}
}
\right)^{\frac{1-n}{n}}
\exp
\left[
\frac{E}{nR}
\left(
\frac{1}{T}
-
\frac{1}{T_{\mathrm{ref}}}
\right)
\right],
\label{eq:dislocation_viscosity}
\end{equation}
where $\eta_{\mathrm{ref}}$ is the reference viscosity,
$\dot{\varepsilon}_{\mathrm{ref}}$ is the reference strain rate,
$n$ is the stress exponent,
$E$ is the activation energy,
$R$ is the gas constant,
$T_{\mathrm{ref}}$ is the reference temperature,
and $\dot{\varepsilon}_{II}$ is the second invariant of the strain-rate tensor.

We use the base-10 logarithm of the reference viscosity as the viscosity parameter,
\begin{equation}
A
=
\log_{10}
\left(
\eta_{\mathrm{ref}}/1~\mathrm{Pa\,s}
\right),
\end{equation}
or equivalently,
$\eta_{\mathrm{ref}}=10^{A}~\mathrm{Pa\,s}$.

The second invariant of the strain-rate tensor is defined as
\begin{equation}
\dot{\varepsilon}_{II}
=
\left(
\frac{1}{2}
\dot{\boldsymbol{\varepsilon}}
:
\dot{\boldsymbol{\varepsilon}}
\right)^{1/2}.
\label{eq:strain_rate_invariant}
\end{equation}

The effective viscosity associated with plastic yielding is defined as
\begin{equation}
\eta_{\mathrm{plas}}
=
\frac{\sigma_y}
{2\dot{\varepsilon}_{II}},
\label{eq:plastic_viscosity}
\end{equation}
where $\sigma_y$ is the yield stress.
When both dislocation creep and plastic yielding are considered,
the effective viscosity is obtained using their harmonic combination:
\begin{equation}
\eta_{\mathrm{eff}}
=
\left(
\eta_{\mathrm{disl}}^{-1}
+
\eta_{\mathrm{plas}}^{-1}
\right)^{-1}.
\label{eq:effective_viscosity}
\end{equation}

In the numerical implementation, the effective viscosity is further bounded smoothly within
\begin{equation}
10^{18}
\leq
\eta_{\mathrm{eff}}
\leq
10^{24}
\ \mathrm{Pa\,s},
\end{equation}
thereby limiting extreme viscosity contrasts while avoiding the non-differentiable points introduced by hard clipping.

Because the effective viscosity depends on both the temperature field and the strain rate determined by the velocity field,
the discrete Stokes equations at each time step form a nonlinear algebraic system.
Denoting the velocity--pressure state by
\begin{equation}
\mathbf{s}
=
\left(
\mathbf{u},p
\right),
\end{equation}
we write the discrete nonlinear Stokes system in residual form as
\begin{equation}
\mathbf{F}
\left(
\mathbf{s};
\boldsymbol{\theta}
\right)
=
\mathbf{0},
\label{eq:nonlinear_stokes_residual}
\end{equation}
where $\boldsymbol{\theta}$ collectively denotes the input quantities at the current time step, including the temperature field, compositional field, and rheological parameters.

For the main implicit-differentiation experiments, we solve the nonlinear system using a combination of Picard fixed-point iterations and Newton iterations.
A number of Picard iterations are first performed to provide a robust initial state for the subsequent Newton solve.
At the $k$th Picard iteration, the strain rate and effective viscosity are evaluated from the current state $\mathbf{s}^{k}$, and a linear Stokes system is solved with the viscosity held fixed:
\begin{equation}
\mathbf{A}
\left(
\mathbf{s}^{k};
\boldsymbol{\theta}
\right)
\mathbf{s}^{k+1}
=
\mathbf{b}
\left(
\boldsymbol{\theta}
\right),
\label{eq:picard_iteration}
\end{equation}
where $\mathbf{A}$ is the discrete Stokes matrix associated with the current viscosity field, and $\mathbf{b}$ is the right-hand side arising from buoyancy forcing and boundary conditions.

Following the Picard iterations, the nonlinear solve switches to Newton iterations.
The Newton correction $\delta\mathbf{s}^{k}$ is obtained from
\begin{equation}
\mathbf{J}^{k}
\delta\mathbf{s}^{k}
=
-
\mathbf{F}
\left(
\mathbf{s}^{k};
\boldsymbol{\theta}
\right),
\label{eq:newton_linear_system}
\end{equation}
where
\begin{equation}
\mathbf{J}^{k}
=
\frac{\partial\mathbf{F}}
{\partial\mathbf{s}}
\left(
\mathbf{s}^{k};
\boldsymbol{\theta}
\right)
\label{eq:newton_jacobian}
\end{equation}
is the full Jacobian of the discrete nonlinear Stokes residual with respect to the velocity--pressure state.
The state is then updated according to
\begin{equation}
\mathbf{s}^{k+1}
=
\mathbf{s}^{k}
+
\delta\mathbf{s}^{k}.
\label{eq:newton_update}
\end{equation}
Picard iteration is generally more robust when the current state is far from the nonlinear solution,
whereas Newton iteration typically provides faster local convergence once the solution is sufficiently close to convergence.
Combining the two provides a balance between robustness and computational efficiency.

Nonlinear convergence is assessed using the normalized residual
\begin{equation}
r_{\mathrm{nl}}^{k}
=
\frac{
\left\|
\mathbf{F}
\left(
\mathbf{s}^{k};
\boldsymbol{\theta}
\right)
\right\|_2
}{
\left\|
\mathbf{b}
\left(
\boldsymbol{\theta}
\right)
\right\|_2
},
\label{eq:nonlinear_relative_residual}
\end{equation}
and the nonlinear Stokes system at the current time step is considered converged when
\begin{equation}
r_{\mathrm{nl}}^{k}
<
\epsilon_{\mathrm{nl}},
\label{eq:nonlinear_tolerance}
\end{equation}
where $\epsilon_{\mathrm{nl}}$ is the prescribed nonlinear convergence tolerance.

For experiments designed to compare gradient-computation strategies,
we additionally treat a fixed number of Picard iterations as the finite-iteration forward model itself.
In this setting, the forward calculation performs exactly $K$ Picard updates, and reverse-mode differentiation is carried out through the same finite sequence of iterations.
This formulation allows us to compare Picard-based unrolled differentiation with implicit differentiation applied to the converged nonlinear Stokes system.

\subsubsection{Time integration of temperature and composition}

Let $T^n$ and $\mathbf{u}^n$ denote the temperature and velocity fields at time step $n$, respectively.
The temperature field is advanced using a diffusion--advection--diffusion (DAD) operator-splitting scheme:
\begin{equation}
T^{n+1}
=
\mathcal{D}_{\Delta t/2}
\circ
\mathcal{A}_{\Delta t}
\circ
\mathcal{D}_{\Delta t/2}
\left(
T^{n};\mathbf{u}^{n}
\right),
\label{eq:dad_temperature}
\end{equation}
where $\mathcal{A}_{\Delta t}$ denotes the semi-Lagrangian advection operator and
$\mathcal{D}_{\Delta t}$ denotes the implicit thermal-diffusion operator.

For a diffusion substep of duration $\Delta t_d$,
$\mathcal{D}_{\Delta t_d}$ is obtained by solving the implicitly discretized equation
\begin{equation}
\rho_0 c_p
\frac{
T^{*}-T^{\mathrm{in}}
}{
\Delta t_d
}
=
\nabla\cdot
\left(
k\nabla T^{*}
\right),
\label{eq:implicit_temperature_diffusion}
\end{equation}
where $T^{\mathrm{in}}$ and $T^{*}$ denote the temperature fields before and after the diffusion substep, respectively.
The spatial diffusion operator is discretized on the $p$-grid using finite differences and assembled into a sparse linear system, which is solved implicitly.
Dirichlet and Neumann temperature boundary conditions are imposed using boundary ghost cells.
In Eq.~\eqref{eq:dad_temperature}, the diffusion operator is applied over a half time step before and after advection, whereas $\mathcal{A}_{\Delta t}$ advances the temperature field over one full time step between the two diffusion substeps.

The compositional field is advected using the same velocity field as the temperature field.
Because a standard single-step semi-Lagrangian scheme can introduce substantial numerical diffusion when transporting non-diffusive scalars, we use a BFECC (back-and-forth error compensation and correction) semi-Lagrangian advection scheme for the compositional field
\citep{dupont2007back,kim2006advections}.
Denoting this operator by $\mathcal{B}_{\Delta t}^{\mathrm{SL}}$, the compositional update is written as
\begin{equation}
C^{n+1}
=
\mathcal{B}_{\Delta t}^{\mathrm{SL}}
\left(
C^n;\mathbf{u}^{n}
\right).
\label{eq:composition_bfecc_operator}
\end{equation}
In the numerical implementation, $\mathcal{B}_{\Delta t}^{\mathrm{SL}}$ corrects the compositional field using an error estimate obtained from successive forward and backward advection steps, thereby reducing numerical diffusion.
Unless otherwise stated, characteristic backtracking is performed using a second-order Runge--Kutta method, and bilinear interpolation is used for both scalar and velocity fields.
For the bounded compositional variable, $C$ is constrained to satisfy
$0 \leq C \leq 1$ after each advection step.

\subsection{Inverse method}
\label{subsec:inverse_method}

We solve the discrete nonlinear mantle-dynamics inverse problem using gradient-based optimization.
Given a set of inversion variables $\mathbf{q}$, the forward model can be represented as a discrete mapping composed of the time-integration operators, the nonlinear Stokes solver, and the observation operator:
\begin{equation}
\mathbf{d}_{\mathrm{pred}}
=
\mathcal{H}
\left[
\mathcal{M}(\mathbf{q})
\right],
\label{eq:inverse_forward_map}
\end{equation}
where $\mathcal{M}$ denotes the forward operator that maps the model parameters to the discrete state variables,
$\mathcal{H}$ denotes the observation operator that maps the model state into the observation space,
and $\mathbf{d}_{\mathrm{pred}}$ denotes the predicted observations.

The inverse problem is formulated as the minimization of an objective function $\mathcal{J}(\mathbf{q})$:
\begin{equation}
\min_{\mathbf{q}} \; \mathcal{J}(\mathbf{q}) .
\label{eq:inverse_optimization_problem}
\end{equation}
At the $k$th optimization iteration, the model parameters are updated according to
\begin{equation}
\mathbf{q}_{k+1}
=
\mathbf{q}_{k}
+
\alpha_k \mathbf{p}_k,
\label{eq:optimization_update}
\end{equation}
where $\mathbf{p}_k$ is a search direction determined from gradient information and $\alpha_k$ is the step length obtained through a line search.
In this study, we primarily use the limited-memory Broyden--Fletcher--Goldfarb--Shanno (L-BFGS) algorithm to optimize the high-dimensional initial temperature field together with low-dimensional physical parameters.

Gradient evaluation is a central component of the inversion framework.
With the exception of the implicit nonlinear Stokes solve performed at each time step, the temperature diffusion, temperature advection, compositional advection, observation operators, interpolation operators, and regularization terms are all implemented using differentiable tensor operations, allowing their derivatives to be evaluated directly using PyTorch automatic differentiation.
For the nonlinear Stokes solver, we consider two differentiation strategies:
unrolled differentiation through a finite number of nonlinear iterations and implicit differentiation based on the converged discrete residual equations.
The two approaches are described in detail below.

\subsubsection{Automatic differentiation framework}
\label{subsubsec:automatic_differentiation}

Automatic differentiation (AD) evaluates derivatives by systematically applying the chain rule to the sequence of differentiable operations that defines the numerical model
\citep{griewank2008evaluating,baydin2018automatic}.
Consider a discrete computation represented by
\begin{equation}
\mathbf{z}_0=\mathbf{q},
\qquad
\mathbf{z}_i
=
\phi_i
\left(
\mathbf{z}_{\mathrm{pa}(i)}
\right),
\quad
i=1,\ldots,N,
\label{eq:ad_computational_graph}
\end{equation}
where $\mathbf{q}$ denotes the input variables,
$\phi_i$ represents a differentiable operation,
and $\mathbf{z}_{\mathrm{pa}(i)}$ denotes the set of preceding variables on which the $i$th operation depends.
For a scalar objective function
\begin{equation}
\mathcal{J}
=
\ell(\mathbf{z}_N),
\label{eq:ad_objective}
\end{equation}
the derivative with respect to the input variables follows from the chain rule applied through this computational graph.

We primarily use reverse-mode automatic differentiation.
Defining
\begin{equation}
\bar{\mathbf{z}}_i
=
\frac{\partial \mathcal{J}}
{\partial \mathbf{z}_i},
\label{eq:ad_adjoint_variable}
\end{equation}
the reverse sweep propagates gradient information from the scalar objective function toward the input variables.
For an operation $\mathbf{z}_i=\phi_i(\mathbf{z}_{\mathrm{pa}(i)})$, the contribution to a preceding variable $\mathbf{z}_j$ is accumulated as
\begin{equation}
\bar{\mathbf{z}}_j
\mathrel{+}=
\left(
\frac{\partial \phi_i}
{\partial \mathbf{z}_j}
\right)^T
\bar{\mathbf{z}}_i,
\qquad
\mathbf{z}_j \in \mathrm{pa}(i).
\label{eq:ad_reverse_mode}
\end{equation}
Reverse-mode AD evaluates the required derivatives through vector--Jacobian products without explicitly forming the full Jacobian.
This is well suited to the inverse problems considered here, where the objective function is scalar and the number of inversion variables can be large.

Within the discrete time-integration framework adopted here, a single time step can be written abstractly as
\begin{equation}
\mathbf{x}^{n+1}
=
\mathcal{G}^{n}
\left(
\mathbf{x}^{n},
\mathbf{s}^{n},
\mathbf{q}
\right),
\label{eq:explicit_timestep_operator}
\end{equation}
where $\mathbf{x}^{n}$ denotes the explicitly evolving state variables, such as the temperature and compositional fields,
$\mathbf{s}^{n}$ denotes the velocity--pressure state obtained from the Stokes equations,
and $\mathbf{q}$ denotes the inversion variables or model parameters.

\subsubsection{Gradient computation for the nonlinear Stokes solver}
\label{subsubsec:differentiable_stokes}

We employ two strategies to compute gradients through the nonlinear Stokes solver at each time step:
unrolled differentiation through a finite number of nonlinear iterations and implicit differentiation based on the converged discrete equations.
The former differentiates directly through the nonlinear iterations that are actually executed,
whereas the latter treats the sufficiently converged velocity--pressure state as an implicit function defined by the discrete nonlinear equations.

Let the velocity--pressure state at time step $n$ be denoted by
\begin{equation}
\mathbf{s}^{n}
=
\left(
\mathbf{u}^{n},p^{n}
\right),
\end{equation}
and let $\boldsymbol{\theta}^{n}$ collectively denote the temperature field, compositional field, and model parameters on which the Stokes system at that time step depends.
The variables $\boldsymbol{\theta}^{n}$ may themselves depend on the inversion variables $\mathbf{q}$ through the preceding time integration.

In unrolled differentiation, the nonlinear Stokes solve is treated as a fixed finite sequence of iterative updates.
Let a single Picard update be written as
\begin{equation}
\mathbf{s}^{n,k+1}
=
\Phi
\left(
\mathbf{s}^{n,k};
\boldsymbol{\theta}^{n}
\right),
\qquad
k=0,\ldots,K-1,
\label{eq:unrolled_fixed_point}
\end{equation}
such that, after $K$ iterations, the Stokes state actually used by the forward model is
\begin{equation}
\mathbf{s}^{n,K}
=
\Phi^{K}
\left(
\mathbf{s}^{n,0};
\boldsymbol{\theta}^{n}
\right).
\label{eq:unrolled_solution}
\end{equation}

During reverse-mode differentiation, PyTorch propagates gradients through the complete computational graph associated with this finite iteration sequence.
The resulting gradient therefore includes all chain-rule dependencies introduced by the individual Picard updates and by the dependence of
$\mathbf{s}^{n,k}$ and $\boldsymbol{\theta}^{n}$ on the inversion variables $\mathbf{q}$.
The Jacobians of the iteration mapping are not formed explicitly.
Instead, the gradient contributions from successive iterations are accumulated through vector--Jacobian products.

Unrolled differentiation computes the discrete gradient of the finite-iteration mapping $\Phi^{K}$.
When the forward calculation and reverse-mode differentiation use the same sequence of $K$ iterations, the resulting gradient remains consistent with the finite-iteration forward model used to evaluate the objective function, even if the state has not reached strict nonlinear convergence.
The main cost is the need to retain intermediate states from the nonlinear iteration history.
Both memory use and reverse-mode differentiation cost increase with the iteration count $K$.
For simulations with many time steps or deeply converged nonlinear iterations, retaining the complete unrolled computational graph can lead to substantial memory and computational overhead.
Implicit differentiation, in contrast, does not differentiate through the nonlinear iteration history.
Instead, the sufficiently converged Stokes state is treated as an implicit function defined by the discrete nonlinear residual equation
\begin{equation}
\mathbf{F}
\left(
\mathbf{s}^{n};
\boldsymbol{\theta}^{n}
\right)
=
\mathbf{0}.
\label{eq:implicit_stokes_constraint}
\end{equation}
For a discrete system written as
\begin{equation}
\mathbf{A}
\left(
\mathbf{s}^{n},\boldsymbol{\theta}^{n}
\right)
\mathbf{s}^{n}
=
\mathbf{b}
\left(
\boldsymbol{\theta}^{n}
\right),
\end{equation}
the nonlinear residual is defined as
\begin{equation}
\mathbf{F}
\left(
\mathbf{s}^{n};
\boldsymbol{\theta}^{n}
\right)
=
\mathbf{A}
\left(
\mathbf{s}^{n},\boldsymbol{\theta}^{n}
\right)
\mathbf{s}^{n}
-
\mathbf{b}
\left(
\boldsymbol{\theta}^{n}
\right).
\label{eq:stokes_residual_definition}
\end{equation}

We define the full Jacobian of the nonlinear residual with respect to the state variables as
\begin{equation}
\mathbf{J}_{\mathbf{s}}^{n}
=
\frac{\partial\mathbf{F}}
{\partial\mathbf{s}^{n}}.
\label{eq:implicit_full_jacobian}
\end{equation}
This Jacobian contains not only the discrete Stokes operator evaluated with viscosity held fixed,
but also the derivative terms arising from the dependence of viscosity on the velocity field through the strain rate.
In the numerical implementation, we construct a coloring map based on the local sparse dependency structure of the discrete residual
and recover the full sparse Jacobian from compressed automatic-differentiation directional derivatives.
This substantially reduces the number of automatic-differentiation evaluations that would otherwise be required to assemble the Jacobian column by column.

Differentiating Eq.~\eqref{eq:implicit_stokes_constraint} with respect to
$\boldsymbol{\theta}^{n}$ gives
\begin{equation}
\mathbf{J}_{\mathbf{s}}^{n}
\frac{\partial\mathbf{s}^{n}}
{\partial\boldsymbol{\theta}^{n}}
+
\frac{\partial\mathbf{F}}
{\partial\boldsymbol{\theta}^{n}}
=
\mathbf{0}.
\label{eq:implicit_constraint_derivative}
\end{equation}
To avoid explicitly forming the state-sensitivity matrix,
we introduce an adjoint variable $\boldsymbol{\lambda}^{n}$ satisfying
\begin{equation}
\left(
\mathbf{J}_{\mathbf{s}}^{n}
\right)^{T}
\boldsymbol{\lambda}^{n}
=
\left(
\frac{\partial\mathcal{J}}
{\partial\mathbf{s}^{n}}
\right)^{T}.
\label{eq:implicit_adjoint_system}
\end{equation}
The contribution of the Stokes solve to the gradient of the objective function with respect to
$\boldsymbol{\theta}^{n}$ is then
\begin{equation}
\frac{\mathrm{d}\mathcal{J}}
{\mathrm{d}\boldsymbol{\theta}^{n}}
=
\frac{\partial\mathcal{J}}
{\partial\boldsymbol{\theta}^{n}}
-
\left(
\boldsymbol{\lambda}^{n}
\right)^{T}
\frac{\partial\mathbf{F}}
{\partial\boldsymbol{\theta}^{n}}.
\label{eq:implicit_gradient}
\end{equation}
The product
$\left(\boldsymbol{\lambda}^{n}\right)^T
\partial\mathbf{F}/\partial\boldsymbol{\theta}^{n}$
is evaluated using automatic-differentiation vector--Jacobian products,
without explicitly assembling
$\partial\mathbf{F}/\partial\boldsymbol{\theta}^{n}$.

Compared with unrolled differentiation, implicit differentiation does not require the complete nonlinear iteration history of each Stokes solve to be stored.
Its local memory requirement does not increase with the number of nonlinear iterations.
The temperature fields, compositional fields, and converged Stokes states needed for reverse propagation across time steps must still be stored or recomputed.
Implicit differentiation also requires the current state to satisfy Eq.~\eqref{eq:implicit_stokes_constraint} with sufficient accuracy.
If the nonlinear residual remains appreciable, the resulting implicit gradient may become inconsistent with the actual finite-accuracy forward mapping.
We use Taylor tests and inversion experiments with different nonlinear residual tolerances to examine how nonlinear solution accuracy affects implicit-gradient accuracy and optimization stability.

\subsubsection{Sobolev preconditioning of the initial temperature field}
\label{subsubsec:sobolev_preconditioner}

Inversion for the initial temperature field constitutes a high-dimensional field inversion problem, and the corresponding gradients may contain pronounced small-scale oscillations.
These high-frequency gradient components can lead to unstable L-BFGS search directions and may cause the optimization to introduce non-physical small-scale temperature perturbations.
To improve the conditioning of the optimization problem, we apply a Sobolev-type gradient preconditioner to the initial temperature field \citep{li2017towards}.

Let $\delta T_0$ denote the initial temperature perturbation to be inverted for.
In the numerical implementation, the inversion variables are defined only at the interior points of the pressure grid:
\begin{equation}
\mathbf{q}_{T}
=
\left(
\delta T_0
\right)_{\mathcal{I}},
\label{eq:temperature_inversion_variable}
\end{equation}
where $\mathcal{I}$ denotes the set of interior grid points excluding the boundaries.
Boundary temperatures are not treated as independent inversion variables but are determined by the prescribed temperature boundary conditions and the lateral Neumann treatment.

Given the raw gradient $\mathbf{g}_{T}$ obtained during the L-BFGS optimization, we do not use it directly for the model update.
Instead, we solve the discrete Helmholtz-type problem
\begin{equation}
\mathbf{M}_{T}\mathbf{z}_{T}
=
\mathbf{g}_{T},
\label{eq:preconditioner_solve}
\end{equation}
and use the resulting $\mathbf{z}_{T}$ as the preconditioned gradient.
The preconditioning matrix is defined as
\begin{equation}
\mathbf{M}_{T}
=
\mathbf{I}
+
\alpha
\mathbf{K}_{T},
\label{eq:preconditioner_matrix}
\end{equation}
where $\mathbf{I}$ is the identity matrix,
$\mathbf{K}_{T}$ is a finite-difference stiffness-like matrix defined on the $p$-grid,
and $\alpha$ controls the smoothing scale of the preconditioner:
\begin{equation}
\alpha
=
\left(
\frac{L_{\mathrm{smooth}}}{L_0}
\right)^2 ,
\label{eq:preconditioner_alpha}
\end{equation}
where $L_{\mathrm{smooth}}$ is a prescribed physical smoothing length and $L_0$ is the characteristic length scale.
This Sobolev preconditioning is equivalent to applying Helmholtz smoothing to the raw gradient.

When the initial temperature field is jointly inverted with low-dimensional
physical parameters, the preconditioner is applied in block-diagonal form:
\begin{equation}
\mathbf{P}
=
\begin{bmatrix}
\mathbf{M}_{T}^{-1} & & & \\
& \mathbf{I}_{\rho_2} & & \\
& & \mathbf{I}_{A} & \\
& & & \mathbf{I}_{n}
\end{bmatrix},
\label{eq:block_diagonal_preconditioner}
\end{equation}
where the Helmholtz-type preconditioner $\mathbf{M}_{T}^{-1}$ acts only on
the high-dimensional initial-temperature block, while identity operators are
used for the low-dimensional physical parameters.

\subsubsection{Objective function and regularization}
\label{subsec:loss_regularization}

We formulate the inverse problem as a discrete constrained optimization problem.
Given the inversion variables $\mathbf{q}$, the objective function consists of a data-misfit term, a smoothness regularization term for the initial temperature field, and, where applicable, a soft bound penalty:
\begin{equation}
\mathcal{J}(\mathbf{q})
=
\mathcal{J}_{\mathrm{data}}(\mathbf{q})
+
\lambda_T \mathcal{R}_{T_0}(\mathbf{q})
+
\lambda_b \mathcal{R}_{b}(\mathbf{q}),
\label{eq:total_objective}
\end{equation}
where $\mathcal{J}_{\mathrm{data}}$ denotes the observational data misfit,
$\mathcal{R}_{T_0}$ denotes the smoothness regularization of the initial temperature field,
$\mathcal{R}_{b}$ denotes the soft bound penalty,
and $\lambda_T$ and $\lambda_b$ are the corresponding weights.
When no explicit temperature-bound penalty is applied, $\lambda_b$ is set to zero.
Following previous adjoint-based studies of time-dependent mantle convection inversion,
we use both the final-time temperature field and time-dependent surface observations to constrain the initial conditions and model parameters.

The total data-misfit term is written as
\begin{equation}
\mathcal{J}_{\mathrm{data}}
=
w_T \mathcal{J}_{T}
+
w_{v_x}\mathcal{J}_{v_x}
+
w_{\tau}\mathcal{J}_{\tau},
\label{eq:data_loss_total}
\end{equation}
where $\mathcal{J}_{T}$ is the final-time temperature misfit,
$\mathcal{J}_{v_x}$ is the misfit of the surface horizontal-velocity time series,
and $\mathcal{J}_{\tau}$ is the misfit in surface traction or surface normal stress, which is a convenient proxy for topographic loading.
The corresponding weights are $w_T$, $w_{v_x}$, and $w_{\tau}$, respectively.
When a particular type of observation is not included in an inversion experiment, its corresponding weight is set to zero.

The final-time temperature misfit is defined using a normalized squared $L_2$ norm:
\begin{equation}
\mathcal{J}_{T}
=
\frac{
\left\|
T_{\mathrm{pred}}^{N_t}
-
T_{\mathrm{obs}}^{N_t}
\right\|_2^2
}{
\left\|
T_{\mathrm{obs}}^{N_t}
-
T_{\mathrm{prior}}
\right\|_2^2
+
\epsilon
},
\label{eq:temperature_loss}
\end{equation}
where $T_{\mathrm{pred}}^{N_t}$ and $T_{\mathrm{obs}}^{N_t}$ denote the predicted and observed temperature fields at the final time, respectively,
$T_{\mathrm{prior}}$ is the prior initial temperature field,
and $\epsilon$ is a small constant introduced to avoid division by zero.
This normalization effectively measures the temperature misfit relative to the magnitude of the temperature anomaly with respect to the prior field.

The surface horizontal-velocity misfit is defined as
\begin{equation}
\mathcal{J}_{v_x}
=
\frac{
\frac{1}{N_t}
\sum_{n=1}^{N_t}
\left\|
v^{n}_{x,\mathrm{surf,pred}}
-
v^{n}_{x,\mathrm{surf,obs}}
\right\|_2^2
}{
\frac{1}{N_t}
\sum_{n=1}^{N_t}
\left\|
v^{n}_{x,\mathrm{surf,obs}}
\right\|_2^2
+
\epsilon
}.
\label{eq:surface_vx_loss}
\end{equation}

The surface-traction misfit is defined using the same time-averaged normalization:
\begin{equation}
\mathcal{J}_{\tau}
=
\frac{
\frac{1}{N_t}
\sum_{n=1}^{N_t}
\left\|
\tau^{n}_{\mathrm{surf,pred}}
-
\tau^{n}_{\mathrm{surf,obs}}
\right\|_2^2
}{
\frac{1}{N_t}
\sum_{n=1}^{N_t}
\left\|
\tau^{n}_{\mathrm{surf,obs}}
\right\|_2^2
+
\epsilon
}.
\label{eq:surface_traction_loss}
\end{equation}

Regularization of the initial temperature field is introduced to suppress non-physical small-scale oscillations.
Defining the initial temperature perturbation as
\begin{equation}
\delta T_0
=
T_0
-
T_{\mathrm{prior}},
\label{eq:temperature_perturbation}
\end{equation}
we employ a first-order smoothness regularization:
\begin{equation}
\mathcal{R}_{T_0}
=
\frac{1}{2}
\left\|
\nabla \delta T_0
\right\|_2^2.
\label{eq:first_order_smoothness}
\end{equation}
In the discrete implementation, Eq.~\eqref{eq:first_order_smoothness} is evaluated using first-order differences between neighbouring points on the $p$-grid.

For inversion experiments in which the internally represented nondimensional initial temperature
$\widetilde{T}_0$ is required to remain within the physically admissible interval $[0,1]$,
we additionally introduce a soft quadratic bound penalty:
\begin{equation}
\mathcal{R}_{b}
=
\frac{1}{N_T}
\sum_{i=1}^{N_T}
\left[
\max
\left(
0,-\widetilde{T}_{0,i}
\right)^2
+
\max
\left(
0,\widetilde{T}_{0,i}-1
\right)^2
\right],
\label{eq:temperature_bound_penalty}
\end{equation}
where $N_T$ is the number of temperature degrees of freedom included in the penalty.
This term vanishes when the initial temperature remains within the admissible interval and increases quadratically when the temperature exceeds either bound.
The penalty discourages non-physical temperature values without imposing a hard projection on the inversion variables.

In addition, to reduce non-physical influences of the lateral boundary conditions and boundary discretization errors on the surface observations, the three observation points nearest to each lateral boundary are excluded when evaluating the surface-velocity and surface-normal-stress misfits. Because the horizontal grid spacing in these regions varies from \(5\) to \(10~\mathrm{km}\) among the different cases, this corresponds to excluding observation points approximately \(15\) to \(30~\mathrm{km}\) adjacent to each lateral boundary from the calculation of the surface-observation misfit.

\subsubsection{Gradient verification}
\label{subsec:gradient_verification}

To verify that the gradients obtained through automatic differentiation and implicit differentiation are consistent with the discrete forward model,
we perform Taylor tests for the principal inversion variables.
Let $\mathbf{q}$ denote the variable to be tested,
$\delta \mathbf{q}$ a unit perturbation direction,
and $\mathcal{J}(\mathbf{q})$ the objective function.
If the computed gradient $\nabla_{\mathbf{q}}\mathcal{J}$ is correct, then, for a sufficiently small perturbation amplitude $h$,
\begin{equation}
\mathcal{J}(\mathbf{q}+h\delta \mathbf{q})
-
\mathcal{J}(\mathbf{q})
=
h
\nabla_{\mathbf{q}}\mathcal{J}^{T}
\delta \mathbf{q}
+
\mathcal{O}(h^2).
\label{eq:taylor_expansion}
\end{equation}

We then examine the first-order Taylor remainder obtained after subtracting the linear gradient contribution:
\begin{equation}
R_1(h)
=
\left|
\mathcal{J}(\mathbf{q}+h\delta \mathbf{q})
-
\mathcal{J}(\mathbf{q})
-
h
\nabla_{\mathbf{q}}\mathcal{J}^{T}
\delta \mathbf{q}
\right|.
\label{eq:taylor_r1}
\end{equation}
For a correctly implemented gradient, the first-order contribution should be accurately cancelled, and $R_1(h)$ should exhibit second-order convergence with decreasing perturbation amplitude:
\begin{equation}
R_1(h)=\mathcal{O}(h^2).
\label{eq:taylor_r1_rate}
\end{equation}

Taylor tests are performed for both the high-dimensional initial temperature field $T_0$ and the low-dimensional physical parameters.
Second-order convergence of the first-order remainder $R_1$ with respect to $h$ indicates that the gradient for the corresponding inversion variable is consistent with the discrete forward model being differentiated.


\section{Results}
\subsection{Sinking lithospheric drip}
\label{subsec:case1_cold_blob}

The first synthetic experiment examines the gravitational foundering of a localized cold lithospheric root and serves as a simple end-to-end test of the differentiable thermo-mechanical inversion framework.
The computational domain is a two-dimensional square of
$500~\mathrm{km}\times500~\mathrm{km}$, discretized using a uniform grid with horizontal and vertical resolutions of $5~\mathrm{km}$.
Free-slip mechanical boundary conditions are imposed on all boundaries.
For temperature, Dirichlet conditions are prescribed at the top and bottom boundaries with
$T_{\mathrm{s}}=273~\mathrm{K}$ and $T_{\mathrm{m}}=1573~\mathrm{K}$, respectively, while zero-flux Neumann conditions are imposed on the lateral boundaries.

\begin{figure}[htbp]
    \centering
    \includegraphics[width=0.70\textwidth]{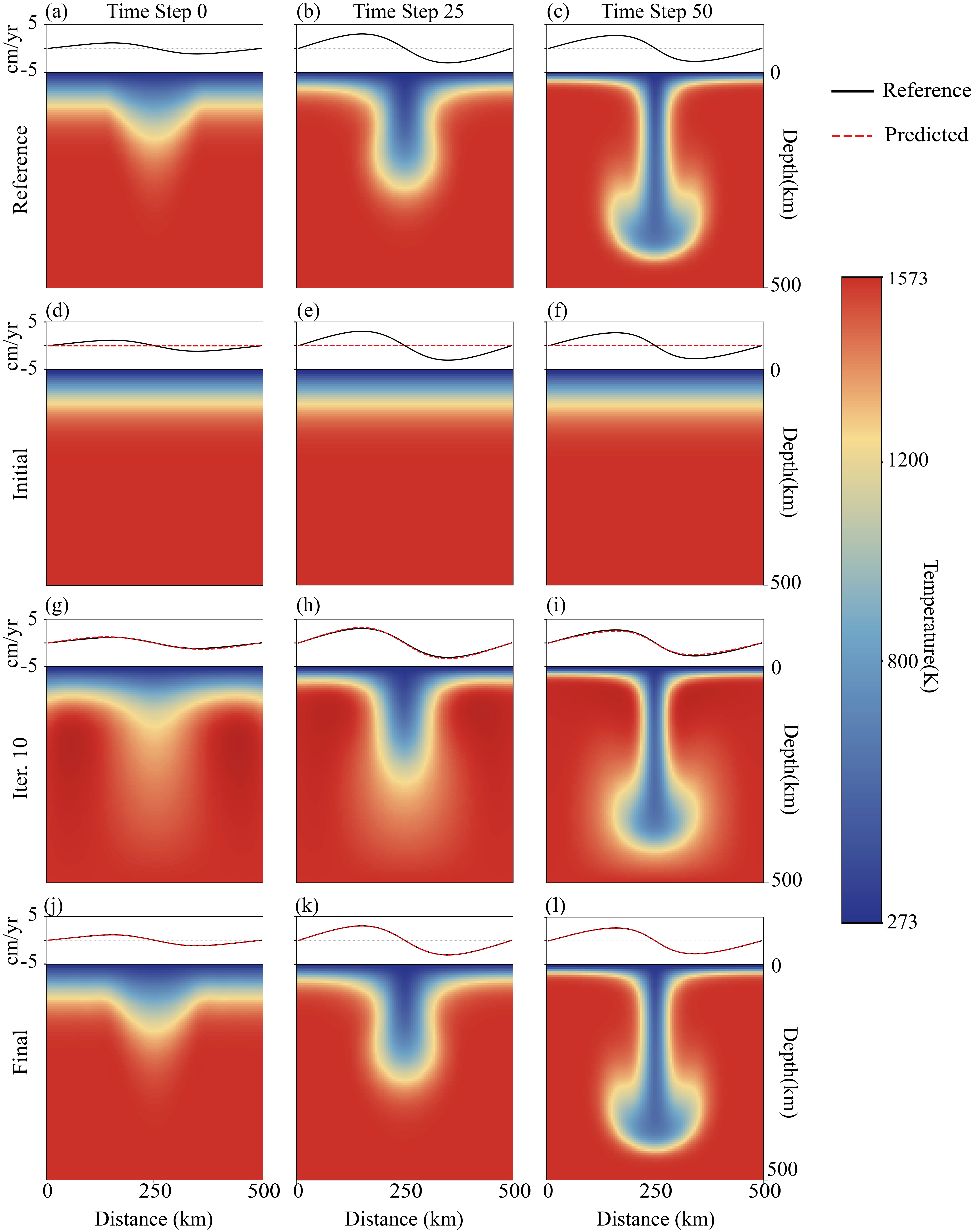}
    \caption{
    Forward and inversion results for Case~1 using Sobolev preconditioning.
    The first row shows the reference model, the second row the forward evolution from the laterally uniform $70~\mathrm{Myr}$ half-space-cooling initial model, the third row the inversion result at iteration~10, and the fourth row the final recovered model.
    The three columns correspond to time steps 0, 25, and 50, respectively.
    Each panel shows the temperature field and the corresponding surface horizontal velocity.
    Black solid and red dashed curves denote the reference and predicted surface velocities, respectively.
    }
    \label{fig:case1_invert}
\end{figure}

The background thermal structure is defined by a laterally uniform half-space-cooling profile corresponding to a plate age of \(70~\mathrm{Myr}\), using a thermal diffusivity of \(\kappa = 10^{-6}~\mathrm{m^2\,s^{-1}}\).
The reference initial condition contains an additional localized thickening of this thermal boundary layer.
To construct a cold lithospheric root that remains smoothly connected to the overlying lithosphere, we locally increase the equivalent half-space-cooling age.
The anomaly is centred at \(x = 250~\mathrm{km}\) and has a horizontal half-width of \(120~\mathrm{km}\).
Within this region, the equivalent cooling age varies smoothly from the background value of \(70~\mathrm{Myr}\) to a maximum value of \(280~\mathrm{Myr}\) at the centre, using a compact cosine-shaped lateral transition.
This produces a half-space-cooling temperature profile that extends progressively deeper toward the centre of the anomaly, forming a smooth downward-convex cold thermal root.
Case~1 employs a spatially and temporally constant viscosity of \(\eta = 10^{21}~\mathrm{Pa\,s}.\)
With constant viscosity, the Stokes problem is linear, and temperature influences the velocity field only through the thermal-buoyancy term.
The reference density is \(\rho_0 = 3300~\mathrm{kg\,m^{-3}}\), and the thermal-expansion coefficient is \(\alpha = 3\times10^{-5}~\mathrm{K^{-1}}\).
No compositional density anomaly, weak zone, plastic yielding, or nonlinear rheology is included in this experiment.
All density variations driving the flow arise solely from the evolving temperature field.
The forward simulation is advanced for \(50\) time steps with a time-step size of \(\Delta t = 2.0\times10^{5}~\mathrm{yr},\)
corresponding to a total model duration of approximately \(10~\mathrm{Myr}\).
The resulting evolution is shown in the first row of Fig.~\ref{fig:case1_invert}.
The negatively buoyant lithospheric root progressively founders into the underlying mantle while undergoing thermal diffusion and deformation.
The associated mantle circulation generates a time-dependent surface horizontal-velocity signal that evolves with the geometry and depth of the sinking thermal structure.

In the inversion experiment, all physical parameters are assumed to be known, and only the initial temperature field is inverted for.
The inversion variables correspond to perturbations of the initial temperature at the interior pressure-grid points.
The upper and lower temperature boundaries remain fixed throughout the inversion, while the lateral boundaries retain the zero-flux condition.
The initial inversion model is the laterally uniform \(70~\mathrm{Myr}\) half-space-cooling lithosphere used to define the background thermal structure.
Because the viscosity is constant, the Stokes equations form a linear system at each time step, and gradients are propagated through the differentiable linear Stokes solve together with the temperature-advection and diffusion operators.
This experiment provides a simple test of the complete time-dependent forward and inversion framework before introducing nonlinear rheology and nonlinear Stokes solution strategies in the following experiments.

\begin{figure}[htbp]
    \centering
    \includegraphics[width=\textwidth]{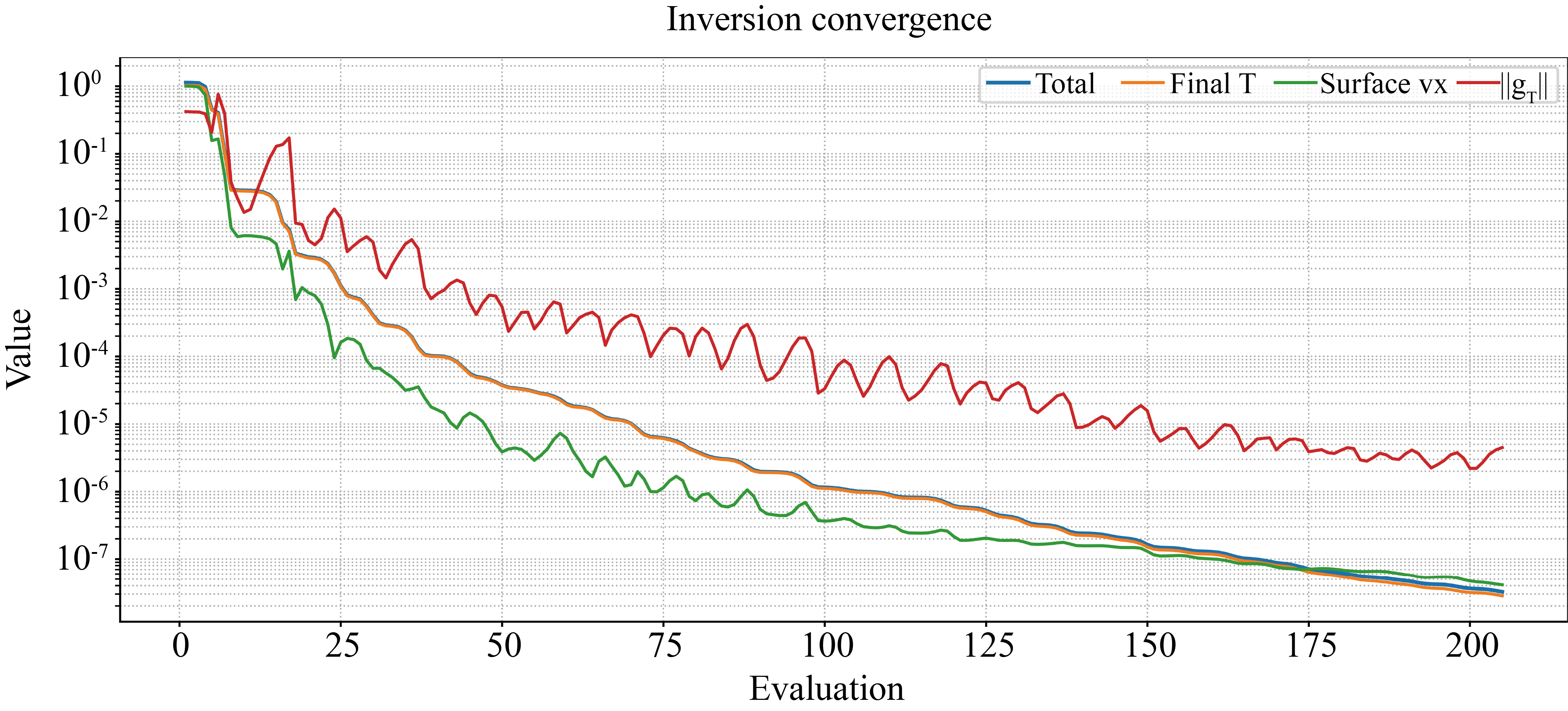}
    \caption{
    Convergence history for Case~1 using Sobolev preconditioning.
    The curves show the total objective function, the final-time temperature misfit,
    the surface-velocity misfit, and the norm of the initial-temperature misfit gradient.
    }
    \label{fig:case1_loss}
\end{figure}

Synthetic observations are generated by forward integration of the reference model and consist of the final-time temperature field and the surface horizontal velocity at every time step.
The objective function combines the final-time temperature misfit and the surface-velocity misfit with weights $w_T=1$ and $w_{v_x}=0.1$, respectively.
No explicit smoothness regularization or soft temperature-bound penalty is applied in this experiment.
Because the initial temperature field is a high-dimensional inversion variable, Sobolev-type preconditioning is applied to its gradient using a physical smoothing length of $50\,\mathrm{km}$.
The preconditioner changes the optimization search direction without modifying the forward equations or the objective function.
The inversion is performed using L-BFGS with a strong-Wolfe line search.
Figure~\ref{fig:case1_loss} shows the evolution of the individual objective-function components during the inversion.
The total objective function, final-time temperature misfit, and surface-velocity misfit all decrease as the cold lithospheric-root structure is recovered.
Figure~\ref{fig:case1_invert} compares the forward evolution of the reference model, the initial inversion model, an intermediate inversion result, and the final recovered model.
Panels~(a)--(c) show the reference evolution, and panels~(d)--(f) show the evolution from the laterally uniform $70\,\mathrm{Myr}$ half-space-cooling initial model.
The initial model does not produce the localized lithospheric root foundering or the associated surface velocity response present in the reference model.
A cold lithospheric root develops during the inversion, and the intermediate model in panels~(g)--(i) already reproduces part of the reference evolution.
The final result in panels~(j)--(l) recovers the main structure of the initial cold root and provides a close match to both the final-time temperature field and the time-dependent surface-velocity response.

\subsection{Comparison of nonlinear Stokes gradient strategies in a thermal subduction model}
\label{subsec:case2_5_subduction}

Case~1 provides a basic test of initial-temperature recovery in a relatively simple thermal model.
We now consider a two-dimensional thermo-mechanical subduction model with a prescribed slab and nonlinear viscosity, where the flow evolution and gradient computation are more complex.
Cases~2--5 use the same physical model, reference initial temperature field, boundary conditions, and numerical discretization.
The only differences among the four experiments are the nonlinear Stokes solution strategy adopted at each time step and the corresponding method used to compute gradients.
Cases~2 and~3 use a fixed number of Picard iterations and apply unrolled differentiation through the finite nonlinear iteration sequence.
Cases~4 and~5 solve the nonlinear Stokes system to a prescribed residual tolerance and compute gradients using implicit differentiation of the converged discrete equations.
The different nonlinear solution strategies produce slightly different velocity--pressure states at each time step, even though the physical setup is the same.
These small differences accumulate over time, so the synthetic data generated in the four cases are not exactly identical.
The comparison focuses on how nonlinear solution accuracy and the choice of differentiated forward mapping affect the inversion of the initial temperature field under the same physical configuration.

\subsubsection{Forward-model setup}
\label{subsubsec:case2_5_forward_setup}

The computational domain is a two-dimensional Cartesian model,
\begin{equation}
L_x \times L_z = 1500~\mathrm{km} \times 660~\mathrm{km}.
\end{equation}
A non-uniform grid is used, with a background resolution of $10~\mathrm{km}$.
To better resolve temperature gradients and velocity variations near the subduction interface, the upper part of the subduction zone is locally refined to a resolution of $5~\mathrm{km}$.
Free-slip mechanical boundary conditions are imposed on all boundaries.
The temperature is fixed at
$T_{\mathrm{top}}=273~\mathrm{K}$
at the top boundary and
$T_{\mathrm{bottom}}=1574~\mathrm{K}$
at the bottom boundary, while zero-flux Neumann conditions are applied at the lateral boundaries.
The initial temperature field is constructed from a half-space cooling model combined with a subducting slab geometry.
A young mid-ocean-ridge region is located on the left side of the model, with the plate age increasing from approximately $0$ to $40~\mathrm{Ma}$ over the leftmost $200~\mathrm{km}$ and remaining at $40~\mathrm{Ma}$ farther to the right.
The slab extends downward and to the right from the vicinity of the trench and is described by a smooth hyperbolic-tangent geometry, with its thermal structure prescribed using a $40~\mathrm{Ma}$ half-space cooling profile.
In the present configuration, the trench is located at approximately $x=550~\mathrm{km}$,
the slab length is approximately $600~\mathrm{km}$,
the thermal slab thickness is approximately $100~\mathrm{km}$,
and the target penetration depth is approximately $300~\mathrm{km}$.
This initial temperature field is defined as the reference model,
$T_{0,\mathrm{true}}$,
and is used to generate the synthetic observations.
A prescribed weak zone is introduced along the subduction interface to represent the mechanically weak region between the slab and the overriding plate.
The geometry of the weak zone remains fixed throughout both the forward and inverse calculations and is not treated as an inversion variable.
The background mantle rheology is temperature- and strain-rate-dependent and includes a plastic-yielding limitation,
whereas the weak zone is represented through a spatially localized viscosity reduction.
The background rheological parameters are
$A=21$,
$n=3$,
and
$E=3.0\times10^{5}~\mathrm{J\,mol^{-1}}$,
with a yield stress of
$\sigma_y=100~\mathrm{MPa}$.
The physical viscosity is smoothly bounded within
\begin{equation}
10^{18}
\leq
\eta
\leq
10^{24}~\mathrm{Pa\,s},
\end{equation}
and the viscosity within the weak zone transitions smoothly towards
$10^{18}~\mathrm{Pa\,s}$.

No compositional anomaly is included in these experiments, and the compositional field is initialized to zero.
The flow is driven entirely by thermal buoyancy.
The reference density is
$\rho=3300~\mathrm{kg\,m^{-3}}$,
and the coefficient of thermal expansion is
$\alpha=3\times10^{-5}~\mathrm{K^{-1}}$.
The forward model is integrated for $30$ time steps, each of duration
$3.0\times10^{5}~\mathrm{yr}$,
corresponding to a total simulation time of approximately
$9~\mathrm{Myr}$.
The forward outputs include the temperature field and surface horizontal velocity at each time step.
The final-time temperature field and the time series of surface horizontal velocity are used in the subsequent inversions.

\subsubsection{Inverse-problem setup}
\label{subsubsec:case2_5_inverse_setup}

In Cases~2--5, the inversion variable is the initial temperature field $T_0$.
In practice, only the initial temperature perturbations at interior pressure-grid points are treated as free optimization variables.
The top and bottom temperature boundaries remain fixed throughout the inversion, while the lateral boundaries are treated using zero-flux conditions.
The initial guess is based on the background half-space cooling temperature field.
This background field preserves the first-order thermal boundary-layer structure and ridge-cooling pattern but does not contain the cold slab anomaly present in the reference model.
The inversion uses the final-time temperature field and the time series of surface velocity to recover the missing subduction-related thermal structure in the initial condition.

Synthetic observations are generated from the corresponding reference forward model and consist of the final-time temperature field and the surface horizontal velocity at each time step.
The objective-function weights are set to
$w_T=1$
for the final-time temperature misfit and
$w_{v_x}=0.1$
for the surface-velocity misfit.
The first-order smoothness regularization applied to the initial temperature field uses
$\lambda_T=10^{-2}$,
and the soft temperature-bound penalty uses
$\lambda_b=1000$.
Sobolev-type preconditioning is applied to the initial-temperature gradient in all four cases to improve the optimization of this high-dimensional variable.
The preconditioner acts only on the initial-temperature perturbations and uses a physical smoothing length of
$L_{\mathrm{smooth}}=50\,\mathrm{km}$.
All inversions are performed using L-BFGS with a strong-Wolfe line search, with a maximum of $200$ L-BFGS iterations.
The rheological parameters, density parameters, weak-zone geometry, and weak-zone viscosity structure remain fixed throughout the inversion.

Cases~2--5 use the same physical configuration, grid, time-step size, boundary conditions, reference initial state, observation definitions, objective-function weights, regularization, and preconditioning.
They differ only in the nonlinear Stokes solution strategy and the way gradients are computed.
Cases~2 and~3 use fixed numbers of Picard iterations with unrolled differentiation:
Case~2 uses $K=5$ iterations and Case~3 uses $K=100$.
Cases~4 and~5 use implicit differentiation, with relative nonlinear Stokes residual tolerances of $10^{-3}$ and $10^{-8}$, respectively.
Because the nonlinear Stokes solves are performed differently, the four cases do not define exactly the same discrete forward mapping.
For the unrolled cases, a larger number of Picard iterations is used at the first time step to obtain a stable initial velocity field, followed by the prescribed fixed iteration counts at subsequent time steps.
In the present model, Case~2 typically leaves nonlinear residuals of about $10^{-2}$--$10^{-3}$, while Case~3 reduces them to about $10^{-7}$--$10^{-8}$.
The four cases span low- and high-accuracy finite-iteration forward models together with implicit-differentiation cases at two different nonlinear convergence levels.

\subsubsection{Inversion results}
\label{subsubsec:case2_5_results}

\begin{figure}[htbp]
    \centering
    \includegraphics[width=0.88\textwidth]{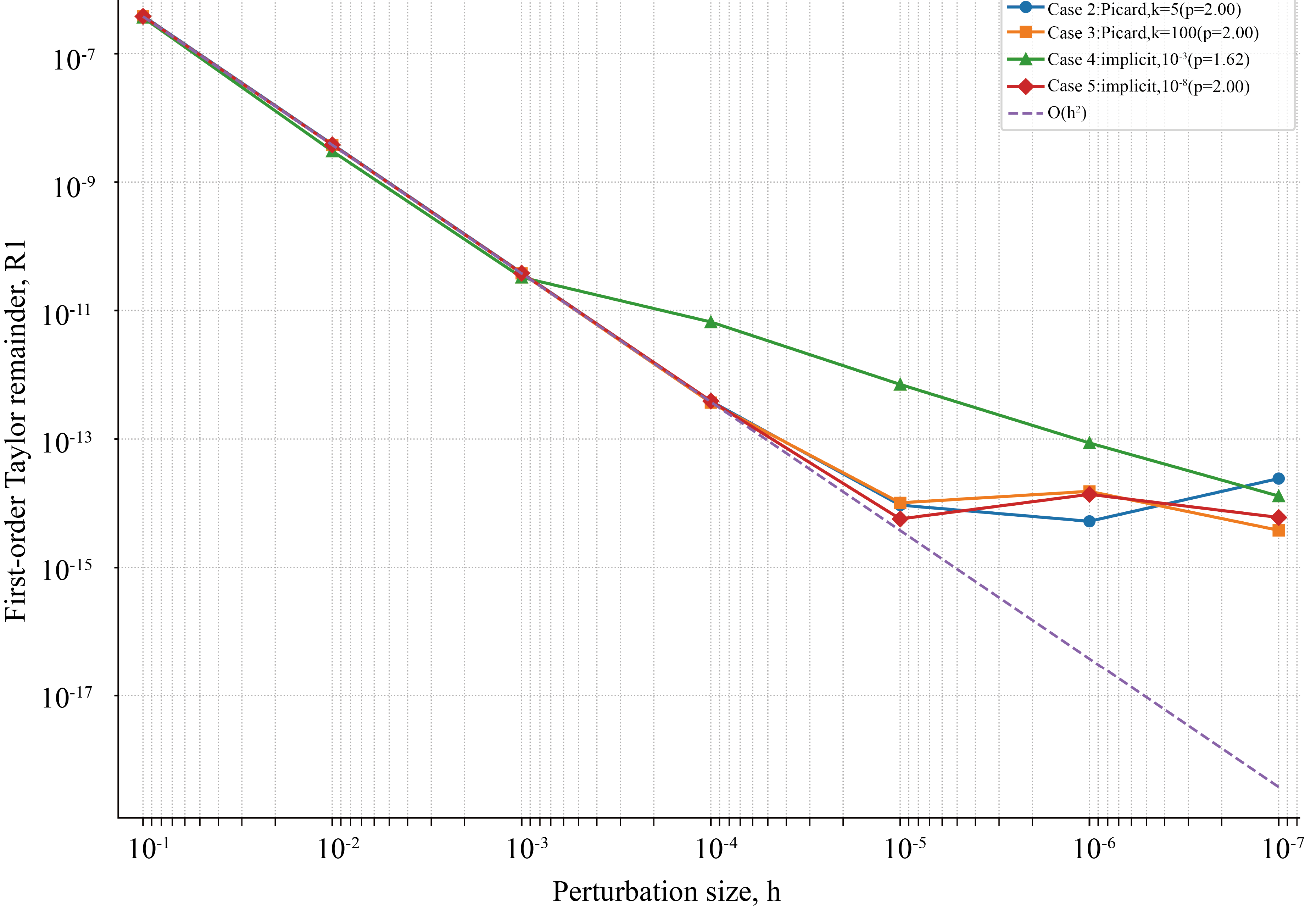}
    \caption{
    Taylor-test results for Cases~2--5.
    The first-order Taylor remainder $R_1$ is shown as a function of perturbation amplitude $h$.
    Here, $K$ denotes the number of fixed-point Picard iterations, while $10^{-3}$ and $10^{-8}$ denote the relative nonlinear Stokes residual tolerances used in the implicit cases.
    The values of $p$ reported in the legend denote the observed convergence orders obtained by fitting $R_1 \propto h^p$ over $h=10^{-1}$--$10^{-4}$.
    The dashed line indicates the reference $O(h^2)$ convergence rate.
    }
    \label{fig:case2_5_taylor}
\end{figure}

Figure~\ref{fig:case2_5_taylor} shows the first-order Taylor remainder $R_1$ as a function of perturbation amplitude $h$.
For Cases~2, 3, and~5, $R_1$ follows approximately second-order convergence over the range $h=10^{-1}$--$10^{-4}$.
At $h=10^{-5}$ and below, the remainder decreases to about $10^{-14}$--$10^{-15}$, close to the limit of double-precision floating-point arithmetic.
Round-off and linear-solver errors then become comparable to the Taylor remainder and cause small deviations from the ideal second-order slope.
The initial-temperature gradients in Cases~2, 3, and~5 are therefore consistent with their corresponding discrete forward models.
Case~4 behaves differently.
Its Taylor remainder is close to second-order over $h=10^{-1}$--$10^{-3}$, but begins to depart from the $O(h^2)$ reference slope when $h\leq10^{-4}$.
Case~4 uses implicit differentiation, but the nonlinear Stokes system is solved only to a relative residual tolerance of $10^{-3}$.
The derivation of the implicit gradient assumes that the current Stokes state satisfies the nonlinear discrete residual equation
$\mathbf{F}(\mathbf{s};\mathbf{m})=\mathbf{0}$
to sufficient accuracy.
When the nonlinear residual remains appreciable, this assumption is not strictly satisfied,
and the gradient obtained through implicit differentiation becomes inconsistent with the actual finite-accuracy forward mapping.
At relatively large perturbation amplitudes, this inconsistency remains small compared with the corresponding change in the objective function,
so that the Taylor test may still exhibit approximately second-order behaviour.
At smaller perturbation amplitudes, however, the nonlinear solution error becomes a dominant contribution,
causing the convergence rate of $R_1$ to deteriorate.
This behaviour shows that the nonlinear tolerance used in Case~4 limits the accuracy of the implicit gradient.

\begin{figure}[htbp]
    \centering
    \includegraphics[width=\textwidth]{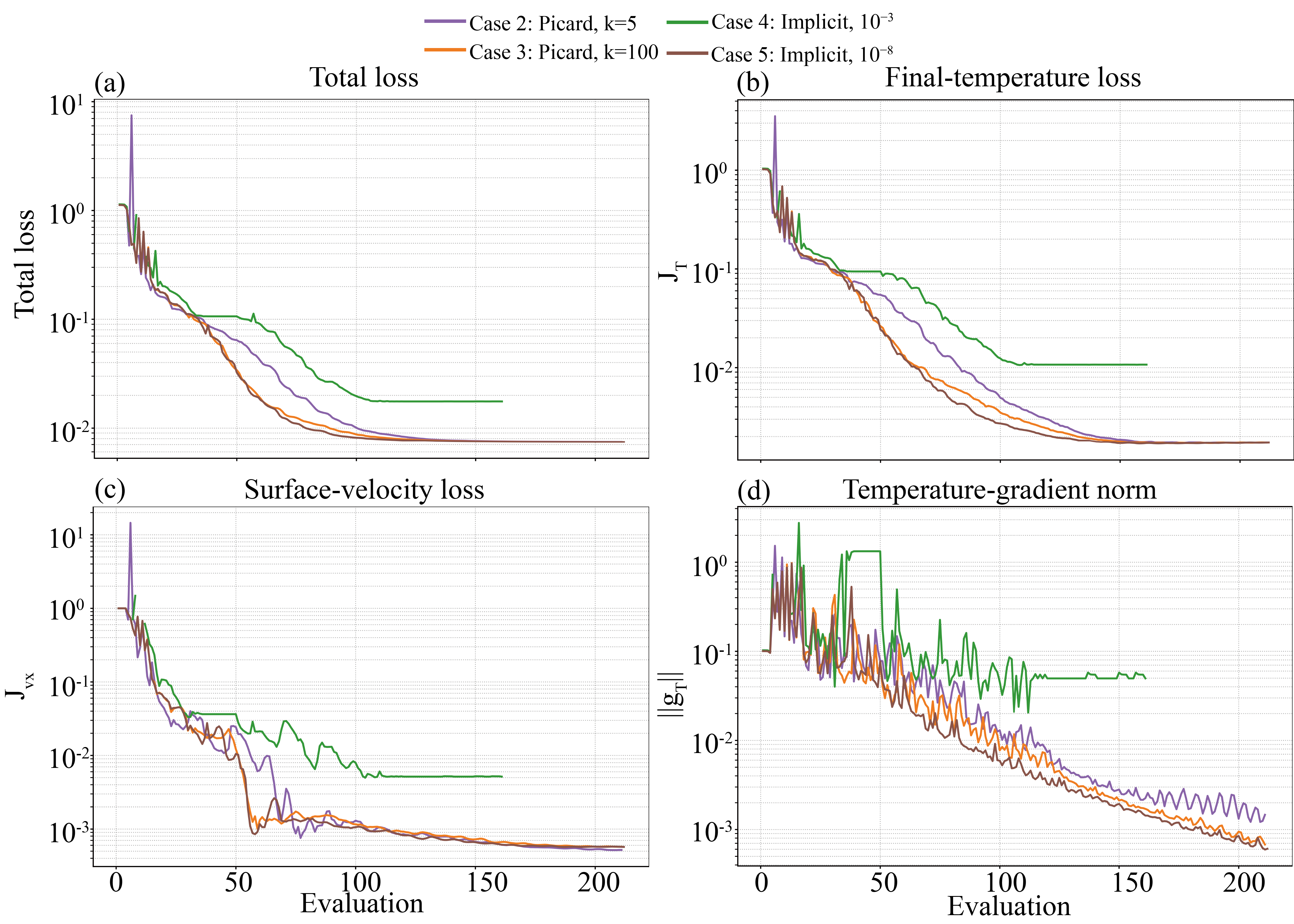}
    \caption{
    Inversion convergence for Cases~2--5.
    The four panels show
    \textbf{(a)} the total objective function,
    \textbf{(b)} the final-time temperature misfit,
    \textbf{(c)} the surface horizontal-velocity misfit, and
    \textbf{(d)} the norm of the initial-temperature misfit gradient,
    as functions of the number of L-BFGS function evaluations.
    }
    \label{fig:case2_5_loss}
\end{figure}

We show the convergence histories of the objective-function components and the norm of the initial-temperature gradient,
$\|\nabla_{T_0}\mathcal{J}\|_2$,
for Cases~2--5 (Fig.~\ref{fig:case2_5_loss}).
The final-time temperature misfit and surface-velocity misfit decrease steadily in Cases~2, 3, and~5,
with reductions of approximately three orders of magnitude within about $200$ function evaluations.
The final data-misfit levels in these three cases are also comparable, and all three cases fit their corresponding synthetic observations well.
Case~4, in contrast, reduces the objective function during the early stages of the inversion,
but the total objective function, final-time temperature misfit, and surface-velocity misfit reach a plateau much earlier.
The final data misfit decreases by only about two orders of magnitude and remains clearly higher than in the other three cases.
The norm of the initial-temperature gradient also remains elevated at later stages, and the optimization does not converge as steadily as in Cases~2, 3, and~5.
This earlier stagnation is consistent with the Taylor-test behaviour of Case~4.
The approximate implicit gradient is still useful during the early stages of the inversion, but the error associated with the finite nonlinear residual becomes increasingly important as the misfit decreases and eventually limits further optimization.

\begin{figure}[htbp]
    \centering
    \includegraphics[width=\textwidth]{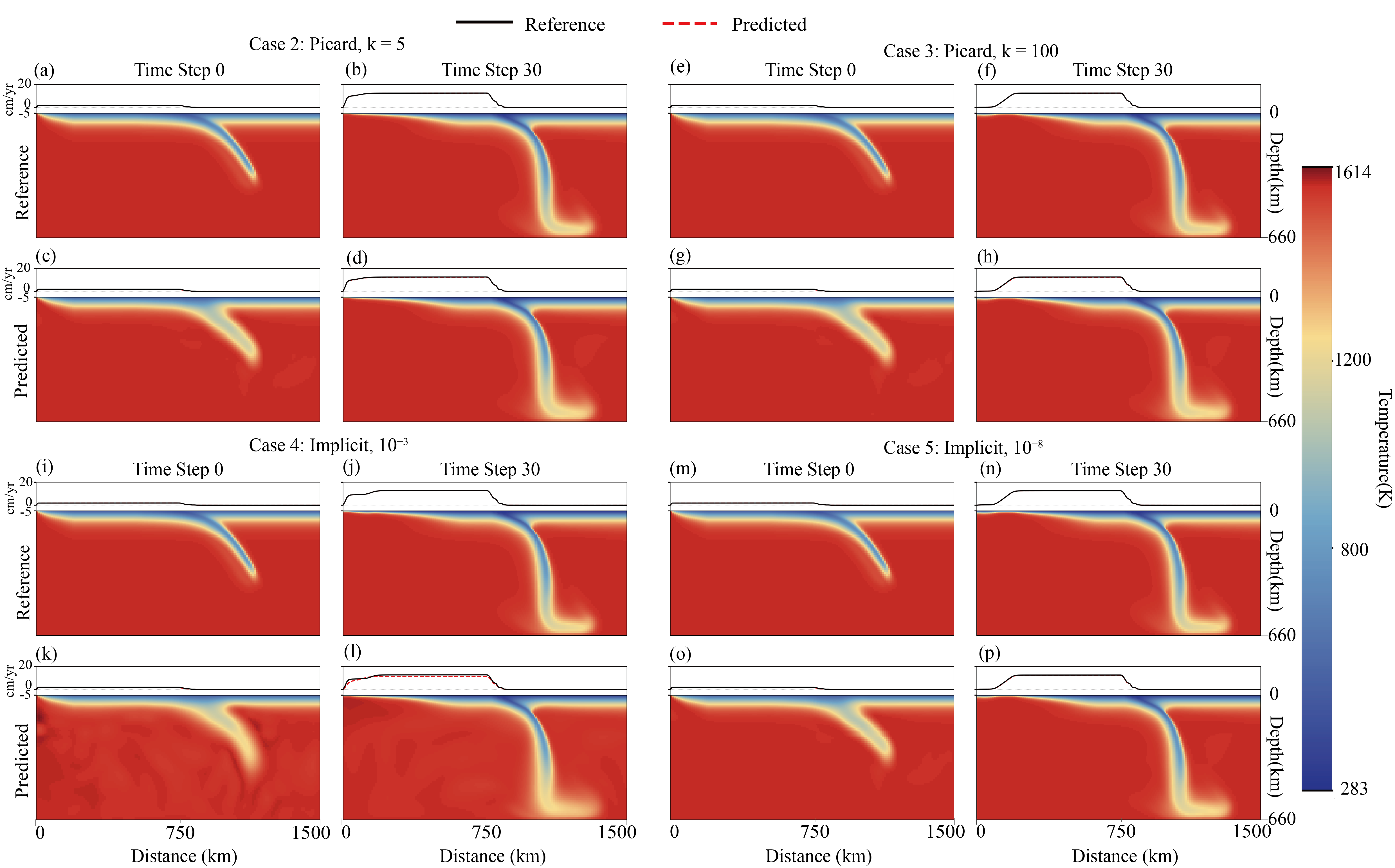}
    \caption{
    Initial-temperature inversion results and corresponding forward evolution for Cases~2--5.
    Panels~(a)--(d), (e)--(h), (i)--(l), and (m)--(p) correspond to Cases~2, 3, 4, and~5, respectively.
    Within each case, the first two panels show the reference model at time steps~0 and~30, and the following two panels show the corresponding inversion results at the same time steps.
    Panels~(a,b), (e,f), (i,j), and~(m,n) show the reference models, whereas panels~(c,d), (g,h), (k,l), and~(o,p) show the inversion results.
    Each panel shows the temperature field together with the corresponding surface horizontal velocity.
    Black solid and red dashed curves denote the reference and predicted surface velocities, respectively.
    }
    \label{fig:case2_5_invert}
\end{figure}

The recovered initial temperature fields show that Cases~2, 3, and~5 reproduce the large-scale cold thermal structure of the subducting slab reasonably well
(Fig.~\ref{fig:case2_5_invert}).
The recovered slab boundaries and small-scale thermal structures are smoother than those in the reference model because of the first-order smoothness regularization and Sobolev preconditioning.
The main location, geometry, and amplitude of the cold slab anomaly are nevertheless recovered well, while some high-wavenumber features are suppressed.
These small-scale differences are progressively reduced by thermal diffusion during the subsequent evolution.
By time step~30, the predicted temperature fields in Cases~2, 3, and~5 are close to their corresponding reference solutions
(Fig.~\ref{fig:case2_5_invert}d, h, p), and the predicted surface horizontal velocities also agree well with the reference results.
This agrees with the approximately three-order-of-magnitude reductions in both the final-time temperature and surface-velocity misfits shown in Fig.~\ref{fig:case2_5_loss}.
Case~4 still recovers the overall cold geometry of the subducting slab, but broader long-wavelength temperature errors remain in the surrounding mantle.
The gradient is sufficient to improve the model during the early stages of the inversion, but these remaining errors are not reduced effectively during later optimization.
Cases~2 and~4 are useful to compare because their nonlinear residuals are of similar magnitude, typically around $10^{-2}$--$10^{-3}$.
Case~2 uses a prescribed sequence of $K=5$ Picard iterations at subsequent time steps and differentiates through the same finite iteration sequence used in the forward calculation.
Its gradient is therefore consistent with the finite-iteration forward mapping used to evaluate the objective function, even though that mapping does not correspond to a fully converged nonlinear Stokes solution.
The optimizer still receives a consistent descent direction, and both the final-time temperature and surface-velocity misfits decrease by about three orders of magnitude.
The recovered initial temperature field is also substantially closer to the reference model than in Case~4.
For a forward model defined by a fixed number of nonlinear iterations, this result shows that unrolled differentiation can preserve consistency between the forward calculation and its gradient even when the nonlinear residual remains relatively large.

Case~5 uses the same implicit-differentiation framework as Case~4, but reduces the relative nonlinear Stokes residual tolerance to $10^{-8}$.
At this tighter tolerance, the nonlinear Stokes solution is sufficiently converged for the implicit gradient to remain consistent with the converged discrete equations.
Both the final-time temperature and surface-velocity misfits decrease by approximately three orders of magnitude, and the recovered initial temperature field is comparable to those obtained in Cases~2 and~3.
The comparison between Cases~4 and~5 shows that the accuracy of implicit differentiation depends strongly on the convergence of the nonlinear Stokes solve.
Case~3 uses a much deeper unrolled Picard solve than Case~2.
The fixed Picard iteration count is increased to $K=100$, reducing the nonlinear residual to approximately $10^{-7}$--$10^{-8}$.
This produces a substantially deeper computational graph and increases both memory use and the cost of reverse-mode differentiation, since the backward pass must propagate through many more nonlinear iterations.
Despite this additional cost, the inversion remains stable.
Both the final-time temperature and surface-velocity misfits decrease by approximately three orders of magnitude, and the recovered initial condition captures the main large-scale cold structure of the subducting slab.
At the model scale considered here, unrolled differentiation thus remains numerically stable even for a relatively deep Picard iteration sequence.

\subsection{Case~6: joint inversion in a thermo-chemical subduction model}
\label{subsec:case6_thermochemical_subduction}

In Cases~2--5, buoyancy is controlled solely by the temperature field, and the inversion variable is limited to the initial temperature field.
These experiments mainly compare different gradient-computation strategies for the nonlinear Stokes solver.
Case~6 extends this setup by introducing a basal thermal boundary layer and a compositional density anomaly, while treating the initial temperature field, the density of the compositional anomaly, and nonlinear rheological parameters as simultaneous inversion variables.
This case tests whether the framework can jointly recover a high-dimensional initial temperature field and several low-dimensional physical parameters in a more complex thermo-chemical model.

Case~6 uses the same basic subduction-model configuration as Cases~2--5, but the fixed bottom temperature is increased to $1874~\mathrm{K}$ to produce a basal thermal boundary layer, and a compositional anomaly is prescribed within several grid layers near the bottom of the model.
The background density is set to
$\rho_0=3300~\mathrm{kg\,m^{-3}}$,
and the density of the anomalous compositional material is
$\rho_2=3350~\mathrm{kg\,m^{-3}}$.
The initial spatial distribution of the compositional field is assumed to be known and is advected by the velocity field during forward integration.
Buoyancy is controlled by both thermal expansion and compositional density contrasts.
The inversion variables are the initial temperature field $T_0$, the density parameter of the compositional anomaly $\rho_2$ (true value: $3350~\mathrm{kg\,m^{-3}}$), the logarithmic reference-viscosity parameter $A$ (true value: $21$), and the stress exponent $n$ (true value: $3$).
The background density $\rho_0$ and the initial spatial distribution of the compositional field remain fixed, so only the density amplitude $\rho_2$ associated with the compositional anomaly is inverted for.
Synthetic observations are generated from the reference forward model and include the final-time temperature field, the time series of surface horizontal velocity, and the time series of surface normal stress.
The surface normal stress provides an additional mechanical constraint related to the dynamic-topography response and helps constrain deep density anomalies and rheological parameters.
The objective-function weights are set to
$w_T=1$
for the final-time temperature misfit,
$w_{v_x}=0.1$
for the surface horizontal-velocity misfit, and
$w_{\tau}=0.1$
for the surface normal-stress misfit.
The first-order smoothness regularization weight for the initial temperature field is
$\lambda_T=10^{-2}$.
Sobolev-type preconditioning is applied to the initial-temperature gradient using a physical smoothing length of
$L_{\mathrm{smooth}}=75~\mathrm{km}$.
The preconditioner acts only on the high-dimensional initial-temperature perturbations; no spatial smoothing is applied to the low-dimensional parameters $\rho_2$, $A$, and $n$.

\begin{figure}[htbp]
    \centering
    \includegraphics[width=\textwidth]{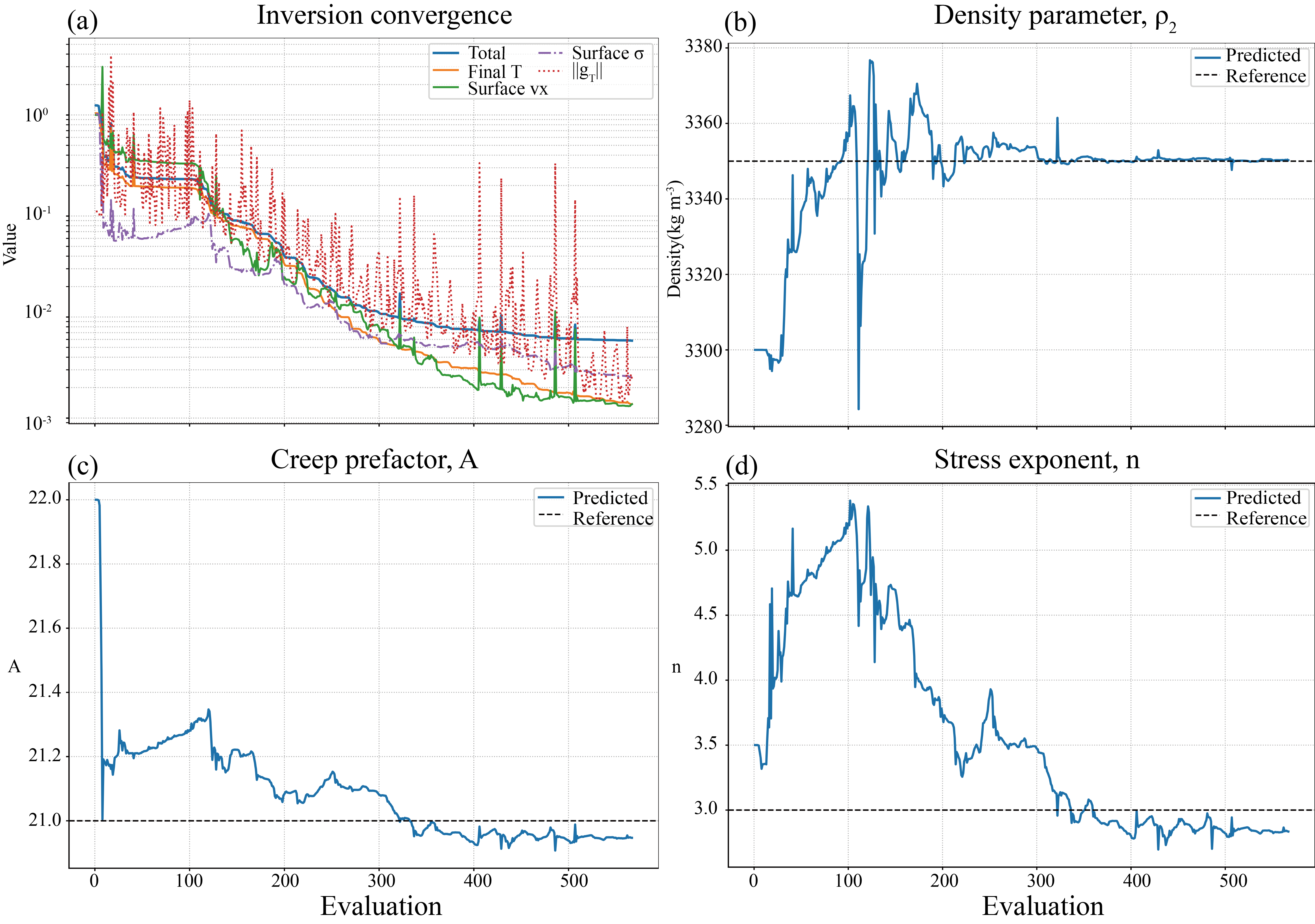}
    \caption{
    Convergence history for Case~6, in which the initial temperature field, the compositional density parameter $\rho_2$, the logarithmic reference-viscosity parameter $A$, and the stress exponent $n$ are inverted jointly.
    \textbf{(a)} shows the evolution of the total objective function, the final-time temperature misfit, the surface horizontal-velocity misfit, the surface normal stress misfit, and the norm of the initial-temperature misfit gradient as functions of the number of L-BFGS function evaluations.
    \textbf{(b)--(d)} show the recovered values of $\rho_2$, $A$, and $n$, respectively, as functions of the number of function evaluations.
    The black dashed lines denote the corresponding true parameter values.
    }
    \label{fig:case6_loss}
\end{figure}

All inversions are performed using the L-BFGS algorithm with a strong-Wolfe line search.
Because this experiment simultaneously inverts for a high-dimensional initial temperature field and several low-dimensional physical parameters, the optimization is allowed to proceed longer than in Cases~2--5.
The maximum number of L-BFGS iterations is set to $500$, and the maximum number of function evaluations is set to $1000$.
At each objective-function evaluation, a complete thermo-mechanical-compositional forward simulation is recomputed for the current values of $T_0$, $\rho_2$, $A$, and $n$, and gradients with respect to these inversion variables are obtained by implicit differentiation.
The convergence history of the objective function and model parameters for Case~6 is shown in Fig.~\ref{fig:case6_loss}.
The convergence curves show that the joint inversion reduces the objective function stably.
The total objective function decreases by approximately two orders of magnitude, whereas the final-time temperature misfit, the surface horizontal-velocity misfit, and the surface normal-stress misfit each decrease by about three orders of magnitude.
The recovered model simultaneously fits the final temperature field and the kinematic and mechanical surface responses.

\begin{figure}[htbp]
    \centering
    \includegraphics[width=\textwidth]{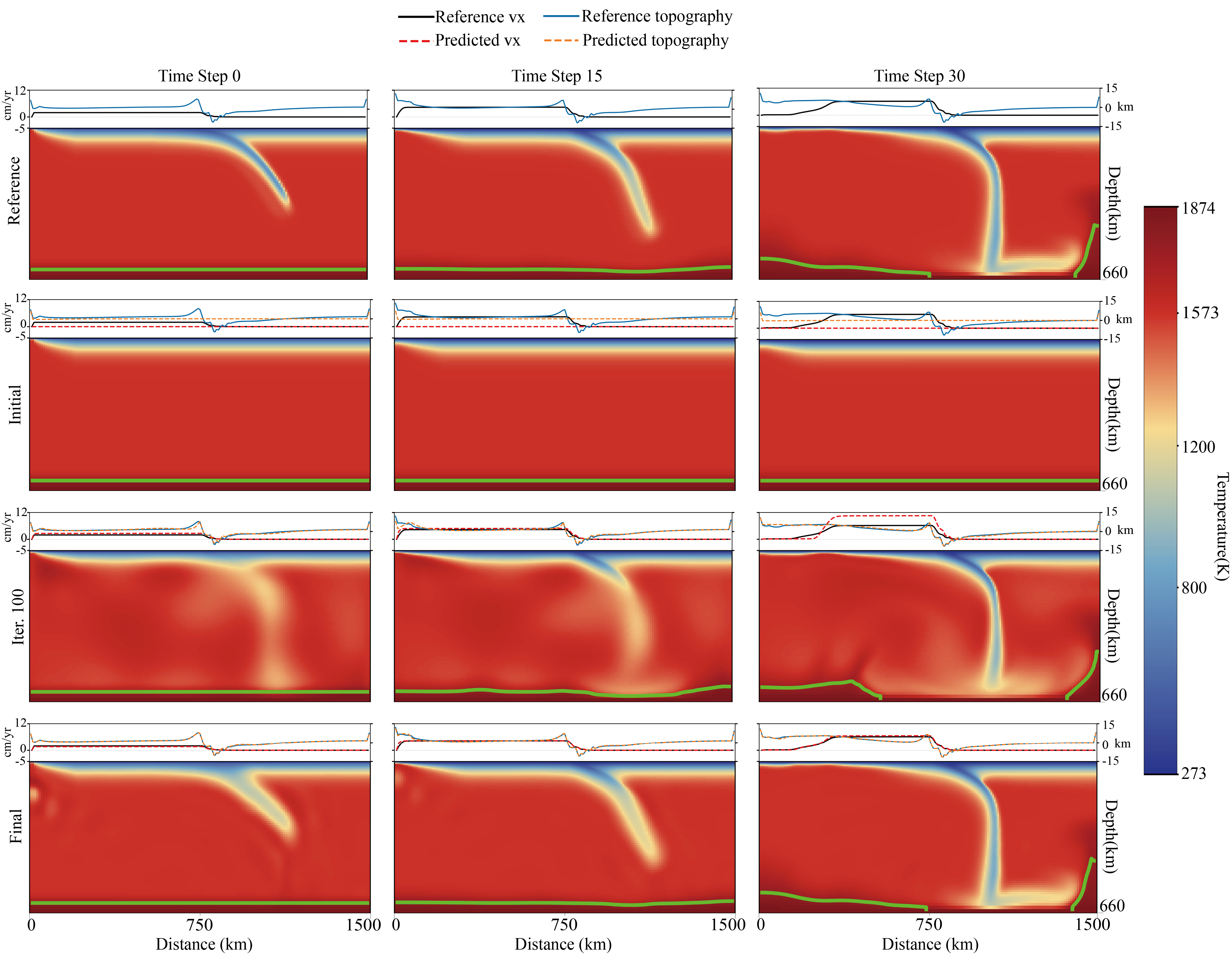}
    \caption{
    Joint inversion results for Case~6 in the subduction model with basal thermo-chemical anomalies.
    The four rows show the reference model, the initial guess, an intermediate inversion result, and the final inversion result, respectively.
    The three columns show the temperature field at the initial time, at time step~15, and at time step~30, respectively.
    The curves above each temperature field show the surface observational responses, where the black solid curves denote the reference-model results and the coloured curves denote the predictions from the corresponding inversion model.
    The green contours mark the compositional interface defined by (C=0.5), separating the two compositional domains.
    The joint inversion recovers the large-scale cold structure of the subducting slab and provides a good fit to both the final-time temperature field and the surface responses after time evolution.
    }
    \label{fig:case6_snapshots}
\end{figure}

The convergence of the low-dimensional parameters is shown in Fig.~\ref{fig:case6_loss}b--d.
The initial guesses for $\rho_2$, $A$, and $n$ all differ from their true values, but all three parameters move toward the reference values during L-BFGS optimization.
The final recovered values are
$\rho_2 \approx 3350\,\mathrm{kg\,m^{-3}}$,
$A \approx 20.95$, and
$n \approx 2.835$.
The compositional density parameter $\rho_2$ is recovered very close to its true value, showing that the combination of surface velocity, surface normal stress, and final-time temperature provides useful constraints on deep compositional buoyancy.
The viscosity parameters $A$ and $n$ also move toward their reference values, indicating that these synthetic observations contain information about the strength of the nonlinear rheology.
Figure~\ref{fig:case6_snapshots} shows the evolution of the temperature field and surface responses during the joint inversion.
The initial guess does not contain the true cold subducting-slab anomaly and fails to reproduce the corresponding thermal structure and surface response of the reference model.
By approximately the 100th function evaluation, part of the large-scale cold structure has been recovered, although the slab geometry and anomaly amplitude still differ from the reference model.
Further optimization recovers the main cold structure of the subducting slab, and the final model agrees reasonably well with the reference solution at time step~30.
Compared with Cases~2--5, Case~6 requires more optimization because the inversion adjusts both a high-dimensional temperature field and several low-dimensional physical parameters.
Some small-scale temperature errors remain in the final model, while the large-scale thermal structure is recovered well.
As in the previous experiments, the remaining high-frequency differences are affected by regularization, Sobolev preconditioning, and thermal diffusion.
The final surface horizontal velocity and surface normal stress also agree well with the reference model, showing that the joint inversion can simultaneously fit the thermal, kinematic, and mechanical observations.

\conclusions[Discussion and conclusions]  

The synthetic experiments show that, under idealized conditions, the differentiable thermo-mechanical framework can recover the main features of the initial temperature field.
For the more complex thermo-chemical model that also includes compositional anomalies, the framework can simultaneously recover a high-dimensional initial temperature field and low-dimensional parameters controlling compositional buoyancy and nonlinear rheology.
When the forward model, observation-generation process, and inversion model are internally consistent, the final-time temperature field, surface horizontal velocity, and surface normal stress can provide useful constraints on the initial thermal structure, compositional density, and rheological parameters.

All inversions considered here are noise-free twin experiments, in which the synthetic observations and inversion calculations use the same governing equations, numerical discretization, computational grid, and boundary conditions.
This provides a useful setting for testing the internal consistency of the forward model, gradient implementation, and optimization procedure, but does not capture the full effects of observational uncertainty, model error, and parameter non-uniqueness that arise in realistic inverse problems
\citep{bunge2003mantle,liu2008simultaneous,li2017towards,hu2024constraining}.
In our previous work on steady-state rheological inversion, we examined how different levels of observational noise affect parameter recovery and optimization stability
\citep{ming2026application}.
The effect of noise on initial-temperature inversion and multiparameter joint inversion has not yet been systematically tested in the present time-dependent framework.
The results presented here should therefore be viewed mainly as a test of what can be recovered under idealized conditions, rather than as evidence that the same level of recovery can be achieved with real Earth observations.
Applying the framework to real geodynamic problems will require consideration of observational noise, data-error covariance, seismic-tomographic resolution, uncertainties in plate reconstructions, boundary conditions, and material properties.
These uncertainties can increase the achievable data misfit and strengthen trade-offs among the initial temperature field, density, and rheological parameters.
Future studies will need to test the inversion under different levels of observational and model uncertainty, use uncertainty-based weighting for different data types, and evaluate the stability and uncertainty of the recovered models.

Previous work has shown that implicit differentiation of nonlinear Stokes systems requires the forward state to satisfy the discrete nonlinear residual equations with sufficient accuracy; otherwise, nonlinear solution errors can contaminate the resulting gradients
\citep{ghelichkhan2024automatic}.
Here we compare unrolled and implicit differentiation over a range of nonlinear residual levels and show that the two methods correspond to different discrete forward mappings.
With unrolled differentiation, the gradient is taken through the finite sequence of nonlinear iterations actually used in the forward calculation.
As long as the forward and reverse calculations use the same fixed iteration sequence, the gradient remains consistent with that finite-iteration model.
The Stokes solution therefore does not need to be fully converged for the gradient to be meaningful with respect to the approximate forward model.
In Case~2, for example, only $K=5$ Picard iterations are used and the nonlinear residual remains around $10^{-2}$--$10^{-3}$, yet the gradient still passes the Taylor test and the inversion remains stable.
The main drawback is memory and computational cost.
Intermediate states from the nonlinear iterations must be retained for the reverse pass, and both storage and differentiation cost increase as more Picard iterations are used.
Case~3 uses $K=100$ iterations and still produces stable gradients, but with a much deeper computational graph and larger memory requirements.
This makes unrolled differentiation a useful option when the forward model is intentionally based on a small fixed number of nonlinear iterations, when an approximate nonlinear solve is acceptable, or when the main goal is rapid testing and algorithm development.

Implicit differentiation takes a different approach. It treats the sufficiently converged Stokes state as the solution of the nonlinear discrete equations and computes sensitivities from this residual-defined solution.
Because the backward pass does not need to retain the full nonlinear iteration history of each Stokes solve, its local memory cost is largely independent of the number of nonlinear iterations. This is useful for larger problems in which many iterations are needed to reach convergence.
The main requirement is that the forward state satisfies
$\mathbf{F}(\mathbf{s};\boldsymbol{\theta})=\mathbf{0}$
to sufficient accuracy.
If the nonlinear solve is stopped too early, the objective function is evaluated using a finite-accuracy Stokes state, while the implicit gradient approximates that of the fully converged residual-defined solution.
The two are then no longer fully consistent.
Cases~4 and~5 show this difference clearly.
With a relative nonlinear residual tolerance of $10^{-3}$, Case~4 shows approximately second-order Taylor convergence only for relatively large perturbations.
At smaller perturbations, the Taylor remainder departs from the expected second-order behaviour, and the inversion eventually reaches a plateau.
When the tolerance is tightened to $10^{-8}$ in Case~5, the Taylor test is recovered and the resulting initial-temperature inversion becomes comparable to the unrolled cases.
This shows that implicit differentiation can work well, but only when the nonlinear Stokes solution is sufficiently accurate.
The trade-off is computational cost.
For high-resolution models, driving the nonlinear residual to a very small value at every time step can be expensive, particularly when full-Jacobian construction, Newton corrections, and accurate linear solves are required.
A fixed number of Picard iterations with unrolled differentiation offers another option when a fully converged nonlinear solve is not necessary: the forward state may remain approximate, but the gradient is still consistent with the finite-iteration model being used.
In practical applications, the preferred strategy depends on the required accuracy of the forward solve, available memory, nonlinear iteration count, and overall model size.

One limitation of the present experiments is that the geometries of the compositional anomaly and weak zone are prescribed rather than inverted.
The inversion only adjusts the initial temperature field, the amplitude of the compositional density contrast, and selected rheological parameters.
If the thermal structure, compositional density, rheology, and geometry are all uncertain, different combinations of these variables may produce similar dynamical responses, leading to strong parameter trade-offs and non-uniqueness
\citep{liu2008simultaneous,li2017towards,ratnaswamy2015adjoint,ming2026application}.
In Case~6, the spatial distribution of the compositional field is assumed to be known, and only its density parameter $\rho_2$ is inverted for.
The position, thickness, and extent of the weak zone are also fixed in the subduction models.
Keeping these geometries fixed makes the inverse problem much easier to handle and lets us focus on the joint recovery of the initial temperature field and a small number of physical parameters.
It also means that the results depend on fairly strong prior information about model geometry.
Some of this geometric information may be available in real applications.
For instance, slab geometry inferred from seismic imaging can help constrain the geometry of weak interfaces, and trench locations from plate reconstructions can help constrain the position of the weak zone.
These geometric choices matter because slab geometry and weak interfaces can strongly affect slab motion and surface observables
\citep{billen2007rheologic,gerya2011,garel2014interaction}.
If the prescribed geometry is wrong, the inversion may partly compensate by changing temperature, density, or rheological parameters.
A model may then fit the observations well while still having an ambiguous physical interpretation.
Directly inverting the full compositional field or weak-zone geometry would introduce many additional degrees of freedom and make the representation and regularization of these structures more difficult.
A more manageable approach would be to describe them with a small number of geometric parameters, such as the extent, thickness, and depth of a compositional anomaly, or the starting position, width, dip, and along-interface extent of a weak zone.
These parameters could be inverted together with the initial temperature field, density, and rheological parameters.
This would relax the assumption of fixed geometry without turning the geometry itself into a fully high-dimensional inversion problem.
Future work can explore this possibility using low-dimensional geometric parameterizations, staged inversion, geometric regularization, or level-set-based approaches.

Overall, this study presents a differentiable thermo-mechanical inversion framework for time-dependent mantle dynamics that supports both high-dimensional initial-condition inversion and joint inversion with selected physical parameters. The comparison between unrolled and implicit differentiation shows that gradient accuracy depends on the discrete forward mapping being differentiated: unrolled differentiation remains consistent with a prescribed finite nonlinear iteration sequence, whereas implicit differentiation requires the nonlinear Stokes state to satisfy the residual equations to sufficient accuracy. In the idealized twin experiments considered here, the framework recovers the main features of the initial thermal structure and jointly constrains compositional density and nonlinear rheological parameters. The framework can be extended in future work to larger-scale models, more realistic observations, and inverse problems that include additional model and geometric uncertainties.




\codedataavailability{
The source code used for the forward simulations and inversion experiments is not publicly available at present and is available from the corresponding author upon reasonable request.}



\appendix

\section{Benchmarks}
\label{sec:benchmarks}

To assess the numerical reliability of the differentiable thermo-mechanical forward framework developed in this study, we consider three benchmarks that test thermal diffusion, thermal convection, and composition-driven slab detachment, respectively.
The Stokes solver itself has already been validated against independent benchmarks in our previous work \citep{ming2026application}.
We therefore do not repeat separate analytical or manufactured-solution tests for the Stokes equations here, but instead focus on the components most directly relevant to the present inversion framework, including thermal diffusion and advection, thermo-mechanical coupling, and compositional advection.

Benchmark~1 considers a two-dimensional thermal-diffusion problem with an analytical solution and is used to verify the thermal-diffusion operator, temperature boundary conditions, and temporal convergence of the time-integration scheme.
Benchmark~2 considers isoviscous thermal convection and is used to test the coupling between the Stokes solver and the temperature advection--diffusion module.
The overall thermal-transport behaviour is evaluated using diagnostic quantities including the Nusselt number and root-mean-square velocity.
Benchmark~3 considers slab detachment and is used to test compositional advection, density-driven buoyancy, nonlinear power-law viscosity, and the dynamical evolution of a system with strong viscosity contrasts.
Together, these three benchmarks cover the principal forward-model components required by the synthetic inversion experiments presented in the main text.

\subsection{Benchmark 1: analytical verification of thermal diffusion}
\label{subsec:benchmark_thermal_diffusion}

We first verify the thermal-diffusion operator and the temporal integration scheme using a two-dimensional diffusion problem with an analytical solution.
This benchmark considers pure thermal diffusion only; the Stokes equations are not solved and the velocity field is set to zero.
The governing equation is
\begin{equation}
\frac{\partial T}{\partial t}
=
\kappa_{\mathrm{nd}}\nabla^2 T ,
\label{eq:benchmark1_diffusion}
\end{equation}
where $\kappa_{\mathrm{nd}}$ is the nondimensional thermal diffusivity.
According to the nondimensionalization adopted in our implementation,
$\kappa_{\mathrm{nd}}=(k/\rho C_p)/\kappa_0$.
For this test, we use
$k=3.0~\mathrm{W\,m^{-1}\,K^{-1}}$,
$\rho C_p=3.3\times10^6~\mathrm{J\,m^{-3}\,K^{-1}}$,
and
$\kappa_0=10^{-6}~\mathrm{m^2\,s^{-1}}$,
giving
$\kappa_{\mathrm{nd}}\approx0.909$.

The computational domain is the same size as that used for the subduction models,
$1500~\mathrm{km}\times660~\mathrm{km}$.
To reduce the influence of spatial discretization errors on the temporal convergence test, a uniform grid spacing of $2~\mathrm{km}$ is used.
Dirichlet conditions with $T=0$ are imposed at the top and bottom boundaries, while zero-flux Neumann conditions are applied at the lateral boundaries.
The corresponding analytical solution is chosen as
\begin{equation}
T(x,z,t)
=
\cos\left(\frac{\pi x}{L_x}\right)
\sin\left(\frac{\pi z}{L_z}\right)
\exp\left[
-\kappa_{\mathrm{nd}}
\left(
\left(\frac{\pi}{L_x}\right)^2
+
\left(\frac{\pi}{L_z}\right)^2
\right)t
\right].
\label{eq:benchmark1_exact_solution}
\end{equation}
This solution satisfies the zero-flux conditions at the lateral boundaries and the zero-temperature conditions at the top and bottom boundaries exactly.

\begin{figure}[htbp]
    \centering
    \includegraphics[width=0.75\textwidth]{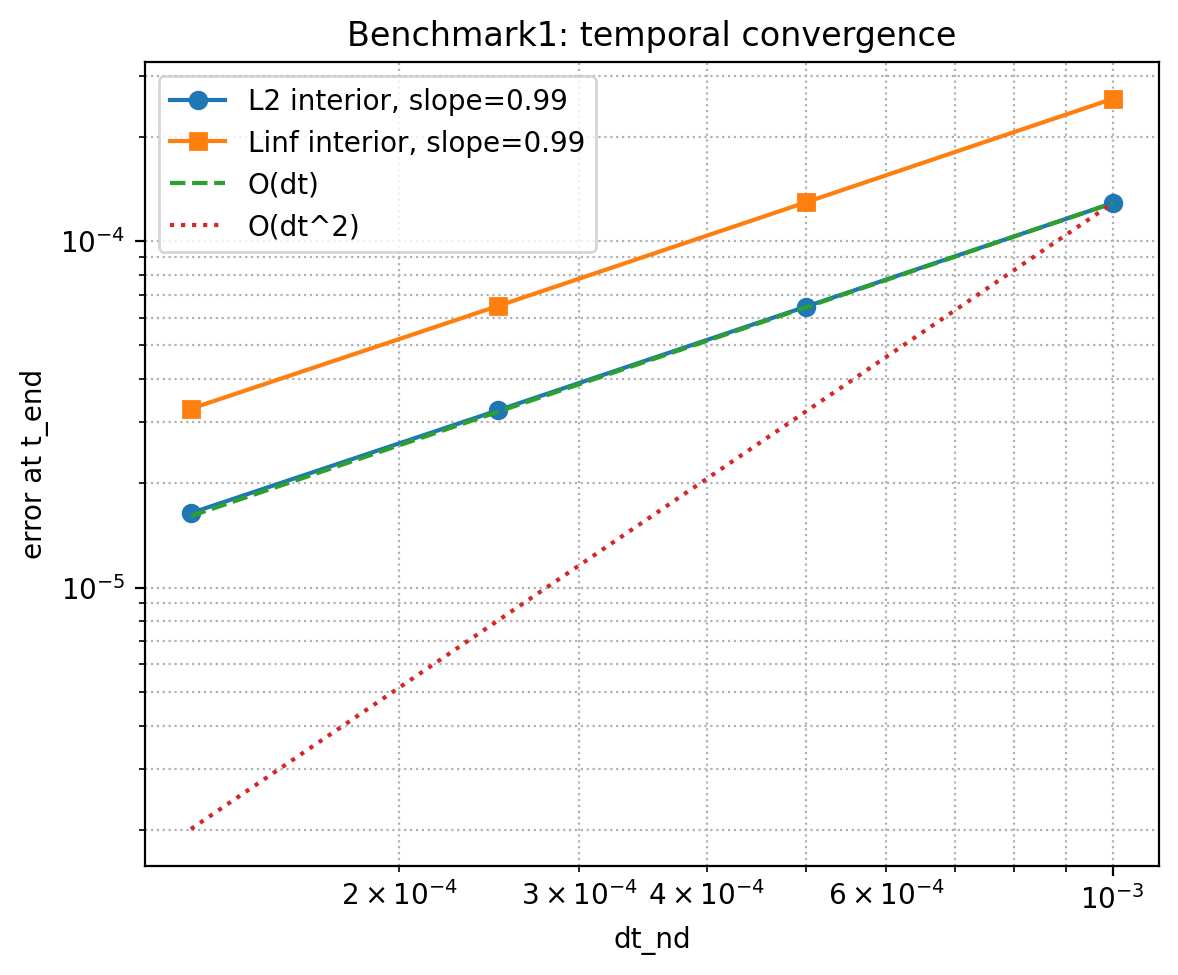}
    \caption{
    Temporal convergence test for Benchmark~1.
    The test uses an analytical solution of the pure thermal-diffusion equation, with a fixed final time and successively decreasing time-step sizes.
    Both the interior $L_2$ and $L_{\infty}$ errors exhibit approximately first-order temporal convergence.
    }
    \label{fig:benchmark1_temporal_convergence}
\end{figure}

For the temporal convergence test, the nondimensional final time is fixed at
$t_{\mathrm{end}}=0.01$.
Simulations are performed using $10$, $20$, $40$, and $80$ time steps, successively halving the time-step size.
Errors are evaluated at the final time, and the $L_2$ and $L_{\infty}$ errors are computed over the interior grid points.
Figure~\ref{fig:benchmark1_temporal_convergence} shows the errors as functions of time-step size.
Both the $L_2$ and $L_{\infty}$ errors decrease approximately linearly with decreasing time-step size, with fitted convergence slopes of approximately $0.99$.
This result demonstrates stable first-order temporal convergence of the thermal-diffusion time integration under the present configuration.

\begin{figure}[htbp]
    \centering
    \includegraphics[width=0.75\textwidth]{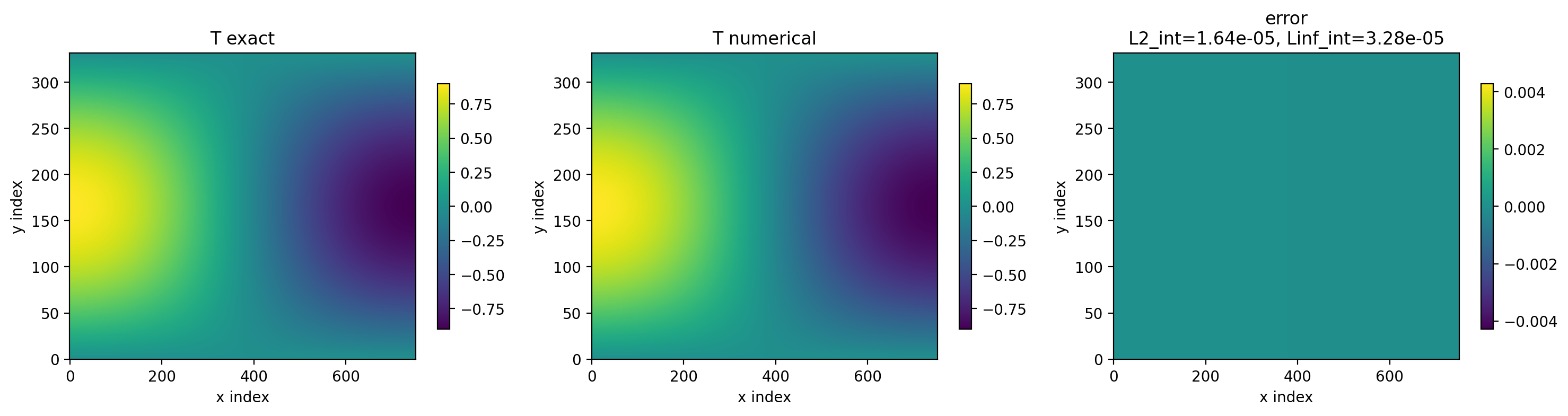}
    \caption{
    Analytical solution, numerical solution, and error field at the final time for the highest-temporal-resolution simulation in Benchmark~1.
    The simulation uses a uniform grid spacing of $2~\mathrm{km}$ and $80$ time steps.
    }
    \label{fig:benchmark1_field}
\end{figure}

Figure~\ref{fig:benchmark1_field} compares the analytical solution, numerical solution, and error field at the final time for the simulation with the highest temporal resolution.
The numerical and analytical solutions show excellent agreement in their spatial structure, and the error remains small throughout the interior of the domain.
For the simulation with $80$ time steps, the interior $L_2$ error is approximately
$1.64\times10^{-5}$,
and the interior $L_{\infty}$ error is approximately
$3.28\times10^{-5}$.
This benchmark confirms that the implemented thermal-diffusion operator and treatment of temperature boundary conditions reproduce the analytical diffusion solution accurately.

\subsection{Benchmark 2: isoviscous thermal convection}
\label{subsec:benchmark_isoviscous_convection}

Following verification of the pure thermal-diffusion operator, we use a two-dimensional isoviscous thermal-convection benchmark to test the overall consistency of the Stokes solver, the temperature advection--diffusion scheme, and the coupled thermo-mechanical workflow.
This benchmark corresponds to the classical two-dimensional isoviscous convection problem in a square box and is widely used for inter-code comparisons of mantle-convection models
\citep{blankenbach1989benchmark}.
The Rayleigh number is set to
$Ra=10^4$,
for which the reference steady-state diagnostic values are
$Nu=4.884409$
and
$V_{\mathrm{rms}}=42.864947$.

The model is defined in a nondimensional unit square.
The governing equations are the incompressible Stokes equations coupled to the temperature advection--diffusion equation.
The viscosity is constant,
$\eta=1$.
Free-slip mechanical boundary conditions are imposed on all boundaries.
The temperature is fixed at
$T=0$
at the top boundary and
$T=1$
at the bottom boundary, while zero-flux Neumann conditions are imposed on the lateral boundaries.

The initial temperature field consists of a linear conductive profile with a small perturbation superimposed.
Specifically, a cosine--sine perturbation with an amplitude of $10^{-2}$ is added to the conductive background profile.
In the buoyancy term, the horizontally averaged temperature at each depth is removed so that only lateral temperature anomalies contribute to the dynamically active buoyancy.
This treatment removes the hydrostatic component that can be absorbed into pressure and ensures that the flow is driven primarily by lateral temperature variations.

The thermal-convection model is integrated using the same Stokes solver and temperature time integrator employed in the main forward framework.
At each time step, the isoviscous Stokes system is first solved using the current temperature field, after which the temperature is advanced using the same advection--diffusion scheme as in the main model.
The time-step size is
$\Delta t=10^{-4}$.
The calculation is continued until both the Nusselt number and root-mean-square velocity reach a steady state.
Steady state is defined by relative changes in both
$Nu$
and
$V_{\mathrm{rms}}$
remaining below
$10^{-5}$
over several consecutive diagnostic intervals.

Tests are performed at four resolutions,
$N=48$, $64$, $96$, and $128$.
Table~\ref{tab:benchmark_isoviscous_convection} lists the steady-state Nusselt numbers and root-mean-square velocities obtained at each resolution.
The Nusselt numbers calculated at the top and bottom boundaries are nearly identical, indicating that the heat flux through the two boundaries is well balanced.
With increasing resolution, the mean Nusselt number increases from
$4.8043$
to
$4.8560$,
progressively approaching the reference value of
$4.884409$.
Similarly,
$V_{\mathrm{rms}}$
decreases from
$44.7891$
to
$43.3785$,
approaching the reference value of
$42.864947$.
At the highest resolution of $128^2$, the relative errors in
$Nu$
and
$V_{\mathrm{rms}}$
are approximately
$0.58\%$
and
$1.20\%$,
respectively.
These results demonstrate that the present implementation reasonably reproduces the steady-state convection structure, overall heat-transport efficiency, and characteristic velocity scale of the benchmark.

\begin{table}[htbp]
\centering
\caption{
Steady-state diagnostics for the isoviscous thermal-convection benchmark at different numerical resolutions.
The reference values correspond to the classical Case~1a benchmark:
$Nu=4.884409$
and
$V_{\mathrm{rms}}=42.864947$.
}
\label{tab:benchmark_isoviscous_convection}
\begin{tabular}{ccccccc}
\hline
Resolution & Step & Time & $Nu_{\mathrm{top}}$ & $Nu_{\mathrm{bottom}}$ & $V_{\mathrm{rms}}$ & Relative error \\
\hline
$48^2$  & 2630 & 0.263 & 4.8043 & 4.8043 & 44.7891 & $1.64\%$, $4.49\%$ \\
$64^2$  & 2660 & 0.266 & 4.8236 & 4.8236 & 44.2265 & $1.24\%$, $3.18\%$ \\
$96^2$  & 2720 & 0.272 & 4.8447 & 4.8447 & 43.6626 & $0.81\%$, $1.86\%$ \\
$128^2$ & 2850 & 0.285 & 4.8560 & 4.8560 & 43.3785 & $0.58\%$, $1.20\%$ \\
\hline
Reference & -- & -- & 4.884409 & 4.884409 & 42.864947 & -- \\
\hline
\end{tabular}
\end{table}

\subsection{Benchmark 3: slab detachment}
\label{subsec:benchmark_slab_detachment}

Following the thermal-diffusion and isoviscous-convection tests, we further consider a slab-detachment benchmark to assess the performance of the code under conditions involving strong viscosity contrasts, nonlinear power-law rheology, and compositional advection.
The benchmark follows commonly used models of buoyancy-driven viscous slab necking and detachment
\citep{schmalholz2011simple,duretz2012dynamics},
and its dynamical evolution is compared qualitatively with the corresponding slab-detachment benchmark implemented in ASPECT\citep{kronbichler2012high,heister2017high,glerum2018nonlinear}.
Because the present implementation and ASPECT differ in their discretization methods, grid configurations, and treatment of compositional transport,
we do not attempt a strict time-by-time quantitative error comparison.
Instead, we examine whether the principal dynamical characteristics, including slab sinking, necking, strain-rate localization, and the surrounding flow pattern, are reproduced consistently.

The benchmark uses a two-dimensional Cartesian domain of
$1000~\mathrm{km}\times660~\mathrm{km}$
with a spatial resolution of
$10~\mathrm{km}$.
The upper lithosphere has a thickness of
$80~\mathrm{km}$,
and the vertically descending slab has a width of
$80~\mathrm{km}$.
The slab is positioned near the centre of the model and extends approximately
$250~\mathrm{km}$
below the base of the lithosphere.
The slab and upper lithosphere are represented by the compositional field
$C=1$,
whereas the surrounding mantle is represented by
$C=0$.

The temperature equation is not solved in this benchmark; the temperature field is retained only as a placeholder required by the Stokes-solver interface.
The dynamics are instead driven by composition-dependent density and viscosity contrasts.
The mantle density is
$\rho_{\mathrm{m}}=3150~\mathrm{kg\,m^{-3}}$,
whereas the slab and lithosphere density is
$\rho_{\mathrm{s}}=3300~\mathrm{kg\,m^{-3}}$.
The slab therefore has a positive density anomaly relative to the surrounding mantle and sinks under gravity.
No-slip conditions are imposed on the left and right boundaries, while free-slip conditions are applied at the top and bottom boundaries.

The background mantle viscosity is constant and set to
$\eta_{\mathrm{m}}=10^{21}~\mathrm{Pa\,s}$.
The slab and lithosphere follow a strain-rate-dependent power-law viscosity of the form
\[
\eta_{\mathrm{s}}
=
\eta_{0,\mathrm{pl}}
\dot{\varepsilon}_{II}^{1/n-1},
\]
where
$\eta_{0,\mathrm{pl}}=4.75\times10^{11}\ \mathrm{Pa\,s^{1/n}}$
and
$n=4$.
The physical viscosity is bounded within the range
$10^{21}$--$10^{25}~\mathrm{Pa\,s}$.
Within compositional transition zones, the mantle and slab viscosities are combined using geometric averaging to avoid excessively sharp numerical transitions at the boundaries of the high-viscosity material.
The compositional field is advected using the semi-Lagrangian BFECC scheme and is constrained to remain within
$[0,1]$
after each time step.
These settings are consistent with the slab-detachment benchmark implementation used in the present code.

\begin{figure}[htbp]
    \centering
    \includegraphics[width=0.85\textwidth]{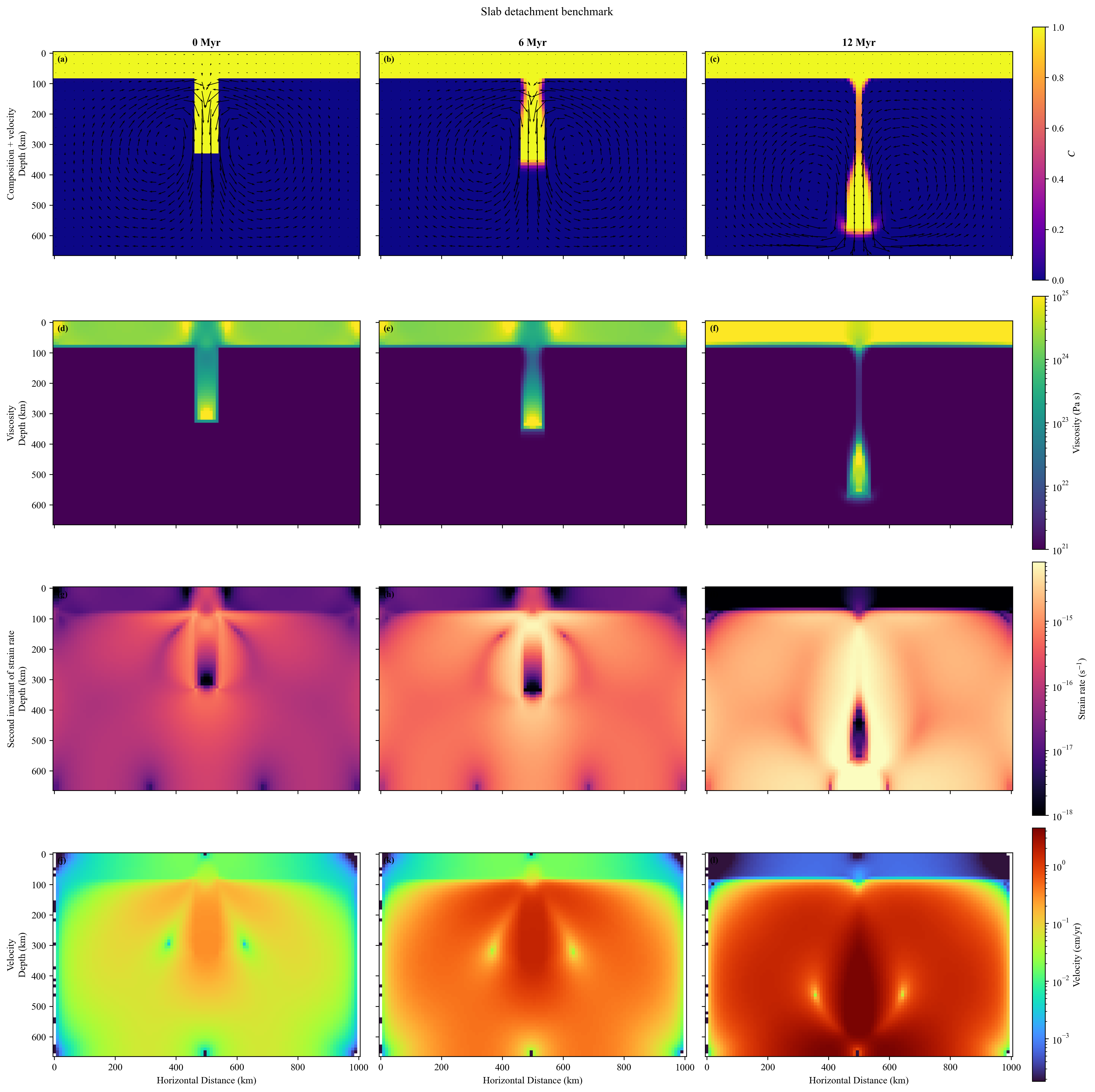}
    \caption{
    Slab-detachment benchmark at $0$, $6$, and $12~\mathrm{Myr}$.
    Each panel shows, from top to bottom, the compositional field with velocity vectors, viscosity field, second invariant of the strain-rate tensor, and velocity magnitude.
    The dense slab sinks under negative buoyancy, while strain rate localizes near the slab neck, producing progressive necking and detachment-like deformation.
    }
    \label{fig:benchmark_slab_detachment}
\end{figure}

The forward time-step size is
$1.0\times10^{5}~\mathrm{yr}$.
We show the model evolution at
$0$, $6$, and $12~\mathrm{Myr}$ (Fig. ~\ref{fig:benchmark_slab_detachment}).
For each time, the figure shows the compositional field with velocity vectors, viscosity field, second invariant of the strain-rate tensor, and velocity magnitude.
Initially, the slab remains attached to the upper lithosphere and forms an approximately vertical, dense, high-viscosity downwelling body.
Driven by its negative buoyancy, the slab sinks and develops strong strain localization near the connection between the slab root and the overlying lithosphere.
As the system evolves, the slab neck progressively narrows and the strain rate increases substantially within the necking region, exhibiting the characteristic behaviour of viscous necking and strain-rate localization.
By
$12~\mathrm{Myr}$,
the main slab body has descended significantly,
a narrow neck has developed between the upper lithosphere and the sinking slab,
and the velocity field exhibits a pair of counter-rotating circulation cells around the descending slab.
These principal dynamical features are qualitatively consistent with the slab sinking, necking, and localized deformation patterns observed in the corresponding ASPECT slab-detachment benchmark.

This benchmark demonstrates that the present Stokes solver can handle strongly localized downwelling driven by compositional density anomalies
and can reproduce physically reasonable slab necking, strain-rate localization, and sinking-induced flow under nonlinear power-law rheology and strong viscosity contrasts.

\section{Gradient verification}
\label{app:gradient_tests}

We verify the inversion gradients using Taylor tests and finite-difference directional-derivative tests.
For an objective function $\mathcal{J}$, a variable $\mathbf{q}$ to be tested, and a unit perturbation direction $\delta \mathbf{q}$, the Taylor expansion is
\begin{equation}
\mathcal{J}(\mathbf{q}+h\delta\mathbf{q})
-
\mathcal{J}(\mathbf{q})
=
h
\nabla_{\mathbf{q}}\mathcal{J}^{T}
\delta\mathbf{q}
+
\mathcal{O}(h^2).
\label{eq:app_taylor_expansion}
\end{equation}
We then define
\begin{equation}
R_0(h)
=
\left|
\mathcal{J}(\mathbf{q}+h\delta\mathbf{q})
-
\mathcal{J}(\mathbf{q})
\right|,
\label{eq:app_taylor_r0}
\end{equation}
and
\begin{equation}
R_1(h)
=
\left|
\mathcal{J}(\mathbf{q}+h\delta\mathbf{q})
-
\mathcal{J}(\mathbf{q})
-
h
\nabla_{\mathbf{q}}\mathcal{J}^{T}
\delta\mathbf{q}
\right|.
\label{eq:app_taylor_r1}
\end{equation}
For a correctly implemented gradient, the expected convergence rates are
\begin{equation}
R_0(h)=\mathcal{O}(h),
\qquad
R_1(h)=\mathcal{O}(h^2).
\label{eq:app_taylor_rates}
\end{equation}

\subsection{Taylor tests for Cases~1 and~6}
\label{app:case1_case6_taylor}

The Taylor-test results for Cases~2--5 are presented in the main text to assess gradient accuracy under different nonlinear Stokes solution and differentiation strategies.
Here, we provide additional Taylor tests for Cases~1 and~6 to verify the gradients in the basic initial-temperature inversion problem and the joint thermo-chemical-rheological inversion problem, respectively.

\begin{figure}[htbp]
    \centering
    \includegraphics[width=0.70\textwidth]{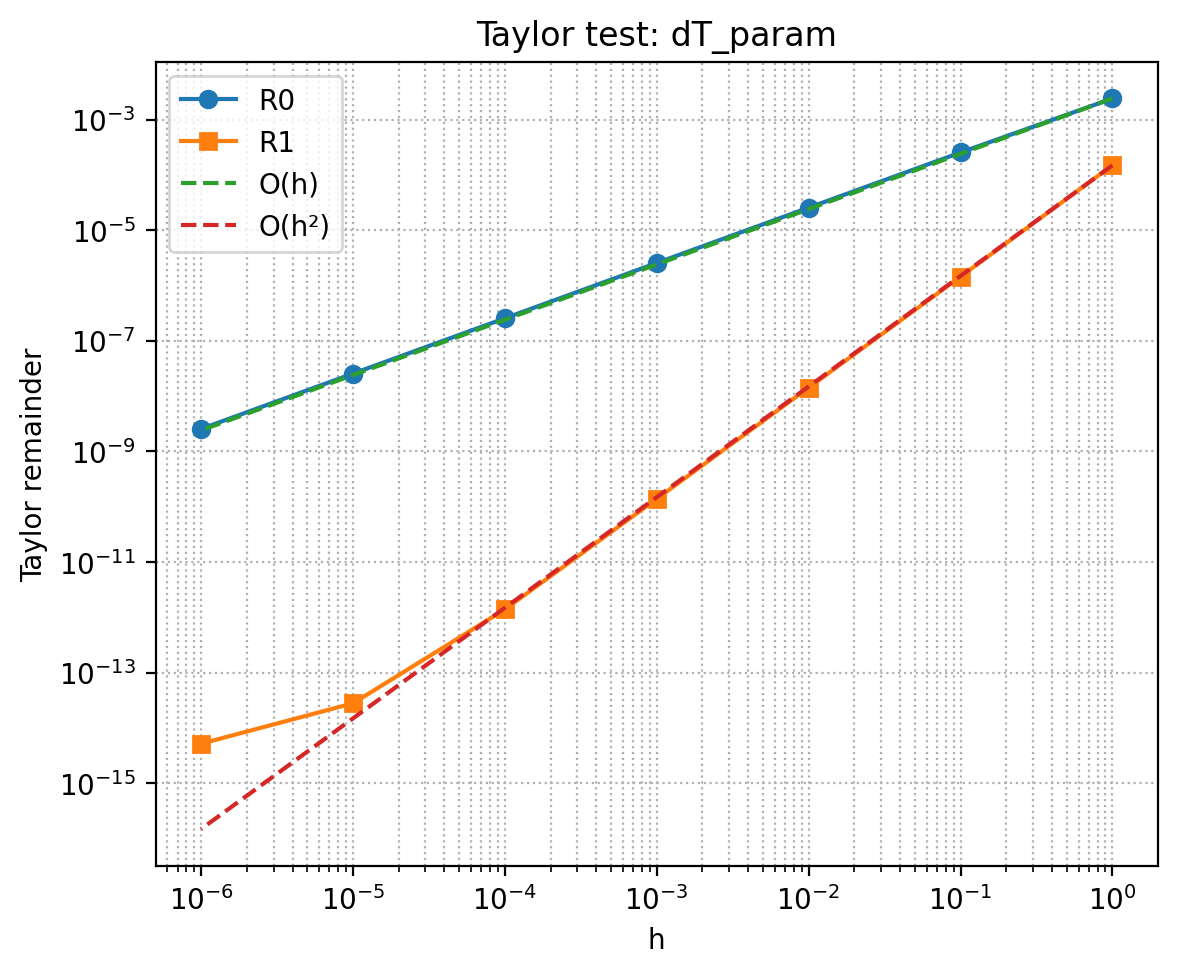}
    \caption{
    Taylor-test results for the initial-temperature inversion variable in Case~1.
    The zeroth-order difference $R_0$ and first-order Taylor remainder $R_1$ are shown as functions of perturbation amplitude $h$.
    The approximately first-order convergence of $R_0$ and second-order convergence of $R_1$ demonstrate that the initial-temperature gradient is consistent with the corresponding discrete forward model.
    At smaller perturbation amplitudes, $R_1$ approaches the limits of double-precision arithmetic, resulting in a slight plateau or deviation from the ideal convergence rate.
    }
    \label{fig:case1_taylor}
\end{figure}

For the isolated sinking cold anomaly in Case~1, a Taylor test is performed with respect to the initial temperature field $T_0$.
As shown in Fig.~\ref{fig:case1_taylor}, the zeroth-order difference $R_0$, for which the linear gradient contribution is not removed, exhibits approximately first-order convergence, whereas the first-order Taylor remainder $R_1$ exhibits second-order convergence.
The second-order convergence of $R_1$ confirms that the gradient with respect to the initial temperature field is consistent with the corresponding discrete forward model.
At smaller perturbation amplitudes, $R_1$ approaches the limit imposed by double-precision floating-point arithmetic, resulting in a slight plateau or departure from the ideal convergence rate.

\begin{figure}[htbp]
    \centering
    \includegraphics[width=0.88\textwidth]{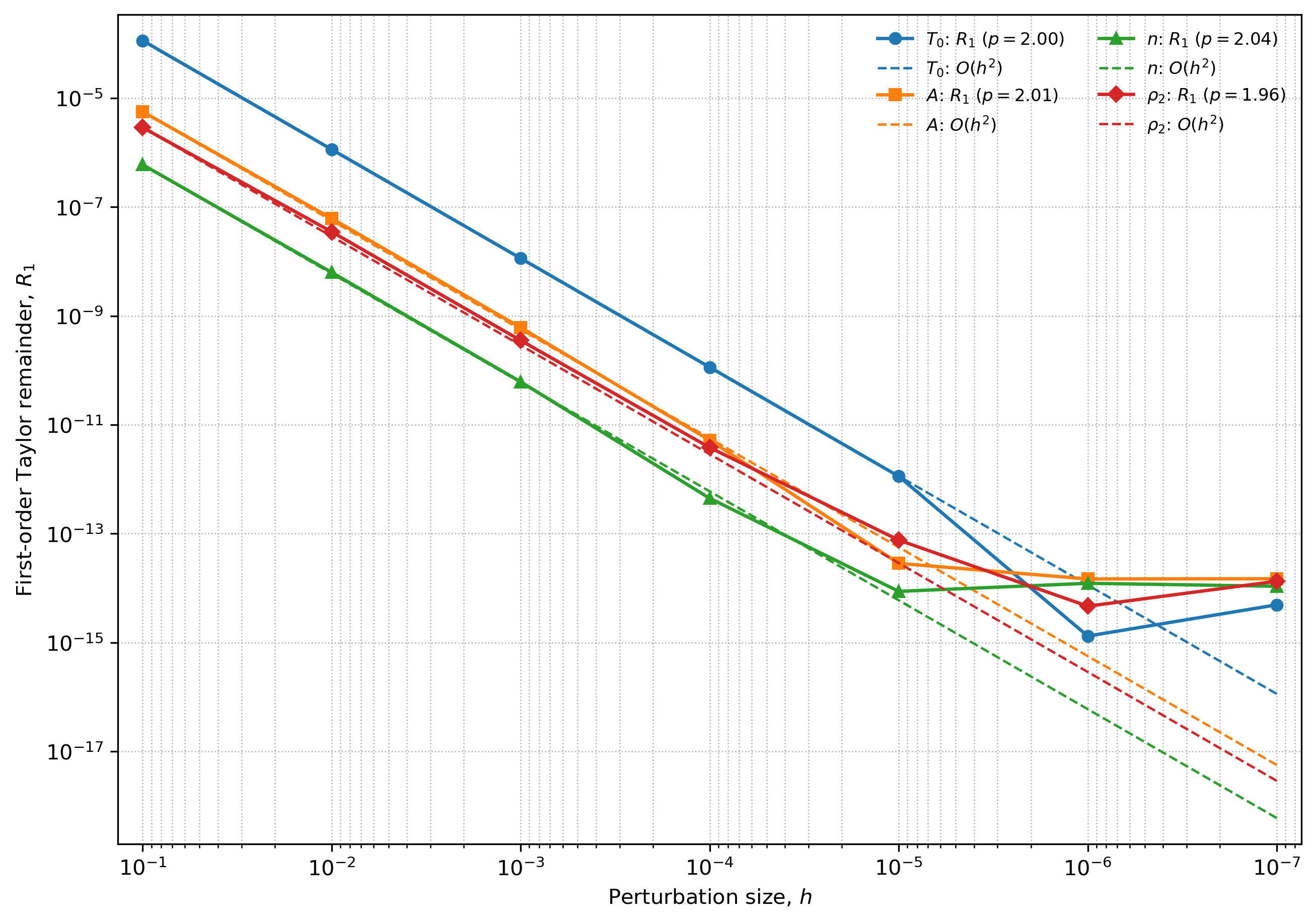}
    \caption{
        Taylor-test results for the different inversion variables in Case~6.
        The first-order Taylor remainder $R_1$ is shown as a function of perturbation amplitude $h$.
        The tested variables are the high-dimensional initial temperature field $T_0$, the logarithmic reference-viscosity parameter $A$, the stress exponent $n$, and the compositional density parameter $\rho_2$.
        The values of $p$ reported in the legend denote the observed convergence orders obtained by fitting $R_1 \propto h^p$ over $h=10^{-1}$--$10^{-4}$.
        The dashed lines indicate the corresponding $O(h^2)$ reference slopes.
        For all four variables, $R_1$ exhibits approximately second-order convergence, demonstrating that the gradients of both the high-dimensional temperature field and the low-dimensional physical parameters are consistent with the discrete forward model.
        At smaller values of $h$, the remainders approach the limits of double-precision arithmetic and show slight deviations from the ideal second-order behaviour.
    }
    \label{fig:case6_taylor}
\end{figure}

For the thermo-chemical subduction model in Case~6, Taylor tests are performed separately for the high-dimensional initial temperature field $T_0$ and the low-dimensional physical parameters $A$, $n$, and $\rho_2$.
As shown in Fig.~\ref{fig:case6_taylor}, the first-order Taylor remainder $R_1$ exhibits approximately second-order convergence for all four inversion variables.
The results confirm that the gradients obtained through implicit and automatic differentiation are consistent with the discrete forward model for both the initial temperature field and the density and rheological parameters.
At smaller perturbation amplitudes, some curves begin to depart from the ideal second-order reference slopes as the Taylor remainders approach the limits of double-precision arithmetic and round-off and linear-solver errors become dominant.

The Taylor tests for Cases~1 and~6 verify the gradients used for the initial-temperature inversion and the joint inversion of $T_0$, $\rho_2$, $A$, and $n$, respectively.
Together with the Taylor tests for Cases~2, 3, and~5 presented in the main text, these tests confirm the consistency of the gradients used in Cases~1, 2, 3, 5, and~6 with their corresponding discrete forward models.
Case~4 provides the contrasting example in which the looser nonlinear Stokes tolerance leads to a measurable inconsistency in the approximate implicit gradient.

\noappendix       




\appendixfigures  

\appendixtables   





\begin{acknowledgements}
This study was supported by the National Key R\&D Program of China through award 2023YFF0806300, the Shenzhen Science and Technology Program through award QNXMA20250701095404006, and the Guangdong Provincial Key Laboratory of Geophysical High-resolution Imaging Technology through award 2022B1212010002.
\end{acknowledgements}



\bibliographystyle{copernicus}
\bibliography{references}

\end{document}